\documentclass{article} 
\usepackage{amsfonts} 

\usepackage{latexsym} 
\usepackage[english]{babel} 
\usepackage{amsmath} 
\usepackage{amstext} 
\usepackage{amsthm} 
\usepackage{xspace} 
  
\newtheorem{theorem}{Theorem}

\newtheorem{corollary}[theorem]{Corollary}

\newtheorem{lemma}[theorem]{Lemma}

\newtheorem{proposition}[theorem]{Proposition} 
\newtheorem{remark}[theorem]{Remark}

\theoremstyle{definition} 
\newtheorem{definition}[theorem]{Definition} 
\input psfig.sty 
 
\begin{document} 
 
\title{Dynamical systems and computable information} 
\author{V.Benci$^1$, C.Bonanno$^2$, S.Galatolo$^{1,}$\footnote{Corresponding author: {\texttt galatolo@mail.dm.unipi.it}} , G.Menconi$^1$, 
M.Virgilio$^3$} \date{} \maketitle  
\small\begin{center} 
{\em $^1$ Dipartimento di Matematica Applicata, Universit\`{a} di 
Pisa, Via Bonanno 26/b, 56125 Pisa, Italy}\\ 
{\em $^2$ Dipartimento di Matematica, Universit\`{a} di 
Pisa, Via Buonarroti 2/a, 56127 Pisa, Italy}\\ 
{\em $^3$ Dipartimento di Fisica, Universit\`{a} di Pisa, Piazza Torricelli 2, 56127 Pisa, Italy } 
\end{center} 
\normalsize 
\begin{abstract} 
We present some new results which relate information to chaotic
dynamics.  In our approach the quantity of information is measured by
the Algorithmic Information Content (Kolmogorov complexity) or by a
sort of computable version of it (Computable Information Content) in
which the information is measured by the use of a suitable universal
data compression algorithm.  We apply these notions to the study of
dynamical systems by considering the asymptotic behavior of the
quantity of information necessary to describe their orbits.  When a
system is ergodic, this method provides an indicator which equals the
Kolmogorov-Sinai entropy almost everywhere.  Moreover, if the entropy
is 0, our method gives new indicators which measure the
unpredictability of the system and allows to classify various kind of
weak chaos. Actually this is the main motivation of this work.  The
behaviour of a zero entropy dynamical system is far to be completely
predictable exept that in particular cases. In fact there are 0
entropy systems which exibit a sort of {\it weak chaos} where the
information necessary to describe the orbit behavior increases with
time more than logarithmically (periodic case) even if less than
linearly (positive entropy case).  Also, we believe that the above
method is useful for the classification of zero entropy time
series. To support this point of view, we show some theoretical and
experimenthal results in specific cases.
\end{abstract} 
\tableofcontents 
\section{Introduction}

In this paper, we present some results on the connections between
information theory and dynamical systems. We analyse the asymptotic behavior
of the quantity of information necessary to describe an orbit of a dynamical
system with a given accuracy. This analysis gives some indicators of
complexity of the orbit itself.

These results have the following features and motivations:

\begin{itemize}
\item  the complexity indicators are defined for a single orbit and can be
estimated numerically; hence they can be used in simulations and in the
analysis of experimental time series;

\item  when the system is ergodic, the orbit complexity equals the
Kolmogorov-Sinai entropy almost surely; thus our method provides a new
characterization of the entropy and an alternative way to compute it;

\item  if the entropy is 0, the asymptotic behavior of the information
provides a measure of the unpredictability of the system and allows to
classify various kind of \textit{weak chaos}. Actually this is the main
motivation of this work.
\end{itemize}

In recent papers (\cite{gal3},\cite{gal2},\cite{Galtesi},\cite{mb})
tools from algorithmic information theory have been used to define and
study some indicators of orbit complexity. These indicators are
invariant up to topological conjugacy. In same special cases the
theory allows to calculate them explicitly and gives a
characterization of various kinds of 0-entropy dynamics. Moreover it
has been proved that there are quantitative relations between these
indicators and the initial condition sensitivity of the system. This
fact shows that information is strongly related to chaos even in the
0-entropy case. The approach of \cite{gal3} makes use of the
Algorithmic Information Content (Kolmogorov complexity) as measure of
the information. Unfortunately the Algorithmic Information Content
(AIC) is not a computable function (see section 2.3) and the related
complexity indicators cannot be used in computer simulations nor in
the analysis of experimental time series.

The aim of this paper is to overcome these difficulties defining orbit
complexity indicators which are suitable for computer experiments. The main
idea consists in replacing the AIC by a \textit{Computable Information
Content} (CIC) which is measured using suitable compression algorithms. We
prove theorems which support the use of this method in the experimental
setting. In particular, we prove that our method gives the same asymptotic
behaviour of the quantity of information when it is measured with AIC in two
important cases: the positive entropy case and the Manneville map which is
an paradigmatic example of intermittent dynamical systems.

Moreover, even when it is not possible to give theoretical estimates, we
have performed some numerical experiments to investigate how the method
works in practice.

\bigskip

The paper is organised as follows.

In Section 2, we recall the main notions of information content for finite
strings and introduce the notion of Computable Information Content whose
definition is based on Compression Algorithms.

In Section 3, we consider infinite strings and we define their complexity as
the time average information content; we prove that the complexity of almost
every string generated by an ergodic information source equals the Shannon
entropy of the source itself.

In Section 4, we consider dynamical systems and, via the symbolic dynamics
method, we apply the results of the previous sections. In particolar, we
prove that the complexity of almost every orbit equals the Kolmogorov-Sinai
entropy, provided that the system is ergodic.

In Section 5, we consider the 0-entropy case (weak chaos) and we introduce
some indicators which are able to detect different kinds of weakly chaotic
dynamics.

In Section 6, we analyse two compression algoritms (LZ77 and CAStoRe) and we
prove some theorem reltive to them which provide a bridge between the
abstract theory and concrete computetions. In particolar we prove that in
the case of the Manneville map the CIC (based on LZ77) provides the same
asymptotic beheaviour than the AIC.

In the final Section (Section 7), we show the results of some numerical
experiment. In the case in which the theory is able to estimate the various
invariants, our empirical results agree with the theoretical ones. In the other
cases, we just show how our method can be applied and we obtain also an
empirical result which, as far as we know, does not have any theoretical
explanation (Casati-Prosen map, \S\ 7.4).

\section{Entropy, information and complexity of finite strings} 
 
The intuitive meaning of \textit{quantity of information} $I(s)$ 
contained in $s$ is the following one: 
 
\begin{center} 
\label{INFO} $I(s)$ \textit{is the length of the smallest binary message 
from which you can reconstruct} $s$. 
\end{center} 
 
Thus, formally  
\begin{equation*} 
I:\mathcal{A}^{\ast }\rightarrow \mathbb{N} 
\end{equation*} 
$I$ is a function from the set of finite strings on a finite alphabet 
$\mathcal{A}$ which takes values in the set of natural numbers. There 
are different notions of information and some of them will be 
discussed here. The first one is due to Shannon. 
 
In his pioneering work, Shannon defined the quantity of information as 
a statistical notion using the tools of probability theory. Thus in 
Shannon framework, the quantity of information which is contained in a 
string depends on its context (\cite{kin}). For example the string 
$^{\prime }pane^{\prime }$ contains a certain information when it is 
considered as a string coming from a given language. For example this 
word contains a certain amount of information in English; the same 
string $^{\prime }pane^{\prime }$ contains much less Shannon 
information when it is considered as a string coming from the Italian 
language because it is much more common (in fact it means 
''bread''). Roughly speaking, the Shannon information of a string $s$ 
is given by 
\begin{equation}  \label{isha} 
I(s)=\log _{2}\frac{1}{p(s)}. 
\end{equation} 
where $p(s)$ denotes the probability of $s$ in a given context. The 
logarithm is taken in base two so that the information can be measured 
in binary digits (bits).\footnote{ From now on, we will use the symbol 
''$\log "$ just for the base 2 logarithm ''$\log _{2}"$ and we will 
denote the natural logarithm by ''$\ln ".$} 
 
If in a language the occurrences of the letters are independent of 
each other, the information carried by each letter is given by 
\begin{equation*} 
I(a_{i})=\log \frac{1}{p_{i}}. 
\end{equation*} 
where $p_{i}$ is the probability of the letter $a_{i}.$ Then the 
average information of each letter is given by 
\begin{equation} 
h=\sum_{i}p_{i}\log \frac{1}{p_{i}}.  \label{ES} 
\end{equation} 
Shannon called the quantity $h$ entropy for its formal similarity with 
Boltzmann's entropy. 
 
We are interested in giving a definition of quantity of information of
a \textit{single} string independent of the context and of any
probability measure. Of course we will require this definition to be
strictly related to the Shannon entropy when we equip the space of all
the strings with a suitable probability measure.
 
In order to be more precise it is necessary to give some notations and 
definitions. 
 
Let us consider a finite alphabet $\mathcal{A}$ and the set 
$\mathcal{A} ^{\ast }$ of finite strings on $\mathcal{A}$, that is 
$\mathcal{A}^{\ast }=\bigcup_{n=1}^{\infty }\mathcal{A}^{n}$. 
 
Now let 
\begin{equation*} 
F:\mathcal{A}^{\ast }\rightarrow \left\{ 0,1\right\} ^{\ast } 
\end{equation*} 
be an injective function, and set 
\begin{eqnarray*} 
I_{F}(s) &=&\left| F(s)\right|  \\ 
K_{F}(s) &=&\frac{\left| F(s)\right| }{\left| s\right| } 
\end{eqnarray*} 
where $\left| s\right| $ is the length of the string $s.$ 
 
Let us consider the usual shift map 
$\sigma :\mathcal{A}^{ \mathbb{N}}\rightarrow \mathcal{A}^{\mathbb{N}}$ 
defined by 
\begin{equation*} 
\left( \sigma (s)\right) _{i}=s_{i+1}\ . 
\end{equation*} 
For a probability measure $\mu$ on $\mathcal{A}^{\mathbb{N}}$, which 
is invariant with respect to the shift, we denote by $h(\mu )$ the 
well-known Shannon entropy of the measure. 
 
Given a string $\omega \in \mathcal{A}^{\mathbb{N}}$, we will denote 
by $\omega ^{n}\in \mathcal{A}^{n}$ the string which consists of the 
first $n$ digits of $\omega .$ 
 
Now we can give the following definition of information and complexity 
 
\begin{definition}[Information measure] \label{jkjlj}
\label{infmeas} If for any ergodic measure $\mu$ on 
$\mathcal{A}^{\mathbb{N}}$ we have that for $\mu$-almost every 
$\omega \in \mathcal{A}^{\mathbb{N}}$  
\begin{equation} 
\limsup\limits_{n\rightarrow +\infty }K_{F}(\omega ^{n})\ =\ h(\mu )\ , 
\label{cond} 
\end{equation} 
then,  
\begin{itemize} 
\item  $F$ is called \emph{ideal coding} 
 
\item $I_{F}(s)$ is called \emph{information content} of $s$ (with 
respect to $F$) 
 
\item $K_{F}(s)$ is called \emph{complexity} (or compression ratio) 
of $s$ (with respect to $F$). 
\end{itemize} 
\end{definition} 
 
Later we will see that ideal codings exist; by condition (\ref{cond}) 
they are asymptotically equivalent to each other. 
 
This definition is given without assuming recursivity for $F$. Later 
on, when we consider $F$ as a {\it coding procedure}, we will mean 
that $F$ is a recursive function. 
 
In the following we will also see that choosing $F$ in a suitable way, 
it is possible to investigate interesting properties of dynamical 
systems with null Kolmogorov-Sinai entropy.  
 
\subsection{Empirical entropy} 
 
The empirical entropy is a quantity that can be thought to be in the 
middle between Shannon entropy and the pointwise definition of 
complexity. The {\it empirical entropy} of a given string is a 
sequence of numbers $\hat{H}_{l}$ giving statistical measures of the 
average information content of the digits of the string $s$. 
 
Let $s$ be a finite string of length $n$. We now define 
$\hat{H}_{l}(s)$, $ l\geq1$, the $l^{th}$ empirical entropy of $s$. We 
first introduce the empirical frequencies of a word in the string $s$: 
let us consider $w\in \mathcal{A}^{l}$, a string on the alphabet 
$\mathcal{A}$ with length $l$; let $s^{(m_{1},m_{2})} \in 
\mathcal{A}^{m_{2}-m_{1}}$ be the string containing the segment of $s$ 
starting from the $m_{1}$-th digit up to the $ m_{2}$-th digit; let 
\begin{equation*} 
\delta(s^{(i+1,i+l)},w)=\biggl\{ 
\begin{array}{cc} 
1 & if\ s^{(i+1,i+l)}=w \\  
0 & otherwise 
\end{array} 
(\ 0 \leq i \leq n-l).  
\end{equation*} 
 
The relative frequency of $w$ (the number of occurrences of the word 
$w$ divided by the total number of $l$-digit sub words) in $s$ is then 
\begin{equation*} 
P(s,w)=\frac1{n-l+1}\sum^{n-l}_{i=0}\delta(s^{(i+1,i+l)},w).  
\end{equation*} 
 
This can be interpreted as the ``empirical'' probability of $w$ 
relative to the string $s$. Then the $l$-empirical entropy is defined 
by 
\begin{equation} \label{l-empent} 
\hat{H}_{l}(s)=-\frac1{l}\sum_{w\in A^{l}}P(s,w) \log P(s,w).  
\end{equation} 
 
The quantity $l\hat{H}_{l}(s)$ is a statistical measure of the average 
information content of the $l-$digit long substrings of $s$.

\subsection{Computable Information Content}

Let us suppose to have some recursive lossless (reversible) coding procedure 
$Z:\mathcal{A}^*\rightarrow\{0,1\}^*$ (for example, the data compression
algorithms that are in any personal computer). Since the coded string
contains all the information that is necessary to reconstruct the original
string, we can consider the length of the coded string as an approximate
measure of the quantity of information that is contained in the original
string.

If $Z$ is an ideal coding (according to defintion \ref{jkjlj} then, as before, the information content of $s$
with respect to $Z$ is defined as $I_{Z}(s)=|Z(s)|.$

Of course not all the coding procedures are equivalent and give the same
performances, so some care is necessary in the definition of information
content. For this reason we introduce the notion of \emph{optimality} of an
algorithm $Z$, defined by comparing its compression ratio with the empirical
entropy.

An algorithm $Z$ is considered optimal if its compression ratio $|Z(s)|/|s|$
is asymptotically less than or equal to $\hat{H}_{k}(s)$ for each $k$.

\begin{definition}\label{ue!}
\textbf{(Optimality)} A reversible coding algorithm $Z$ is \textit{optimal }%
if $\forall k\in \mathbf{N}$ there is a function $f_{k}$, with $f_{k}(n)=o(n)
$, such that for all finite strings $s$ 
\begin{equation*}\label{optimal}
\frac{|Z(s)|}{|s|}\leq \hat{H}_{k}(s)+\frac{f_{k}(|s|)}{|s|}.
\end{equation*}
\end{definition}

Many data compression algorithms that are used in applications are proved to
be optimal.

\begin{remark}
The universal coding algorithms $LZ77$ and $LZ78$ (\cite{lz77},\cite{lz78})
satisfy Definition \ref{ue!}. For the proof see \cite{manzini}.
\end{remark}

Using the definition above, we are able to define the Computable Information
Content of a string:

\begin{definition}
The Computable Information Content of a string $s$ is an information measure
(in the sense of Def. \ref{jkjlj}) where the ideal coding $F$ is an optimal
compression algorithm.
\end{definition}

\bigskip 

The notion of optimality is not enough if we ask a coding algorithm to be
able to reproduce the rate of convergence of the sequence $\hat{H}_{k}(s)$
as $|s|\rightarrow \infty $ for strings generated by weakly chaotic
dynamical systems, for which $\lim_{|s|\rightarrow \infty }\hat{H}_{k}(s)=0$%
. Indeed, if in the positive entropy systems optimality implies that
asymptotically $\frac{|Z(s)|}{|s|}$ is equivalent to $\hat{H}_{k}(s)$, in
the weakly chaotic systems it may happen that the asymptotic behavior
dominant in the right hand side of equation (\ref{optimal}) is that of the
function $\frac{f_{k}(|s|)}{|s|}$.

For example let us consider the string $0^n1$ and the $LZ78$ algorithm, then 
$\hat{H}_k (0^n1)$ goes like $log(n)/n$ while $LZ78(0^n1)/n$ goes like $%
\frac{n^{1/2}log(n)}{n}$ (see also \cite{ChaSolFra}). This implies that
optimality is not sufficient to have a coding algorithm able to characterize 
$0-$entropy strings according to the rate of convergence of their entropy to 
$0$. For this aim we need an algorithm having the same asymptotic behavior
of the empirical entropy. In this way even in the 0-entropy case our
algorithm will provide a meaningful measure of the information. The following
definition (from \cite{manzini}) is an approach to define optimality of a
compression algorithm for the 0-entropy case.

\begin{definition}[Asymptotic Optimality]
A compression algorithm $Z$ is called \emph{asymptotically optimal} with
respect to $\hat{H}_{k}$ if it is optimal and there is a function $g_{k}$
with $g_{k}(n)=o(n)$ and $\lambda >0$ such that $\forall s$ with $\hat{H}%
_{k}(s)\neq 0$ 
\begin{equation*}
|Z(s)|\leq \lambda |s|\hat{H}_{k}(s)+g_{k}(|Z(s)|).
\end{equation*}
\end{definition}

It is not trivial to construct an asymptotically optimal algorithm. For
instance the well known Lempel-Ziv compression algorithms are not
asymptotically optimal. $LZ78$ is not asymptotically optimal even with
respect to $\hat{H}_{1}$ (\cite{manzini}). In \cite{manzini} some examples
are described of algorithms (LZ78 with RLE and LZ77) which are
asymptotically optimal with respect to $\hat{H}_{1}$. But these examples are
not asymptotically optimal for each $\hat{H}_{k}$ with $k\geq 2$. The
asymptotic optimality of LZ77 with respect to $\hat{H}_{1}$ (Theorem \ref
{teomanz}) however is enough to prove (see Section \ref{man77}, Theorem \ref
{41}) that LZ77 can estimate correctly the information coming from an
important example of weak chaos: the Manneville map.

The set of asymptotically optimal compression algorithms with respect to
each $\hat{H}_{k}$ is not empty. In \cite{Galtesi} an exam\-ple is given of
a compression algorithm that is asymptotically optimal for each $\hat{H}_{k}$%
. The algorithm is similar to the Kolmogorov frequency coding algorithm
which is also used in \cite{brud}. This compression algorithm is not of
practical use because of its computational complexity.

To our knowledge the problem of finding a fast asymptotically optimal
compression algorithm is still open.

\subsection{Algorithmic Information Content} 
 
One of the most important information function is the Algorithmic 
Information Content ($AIC$). In order to define it, it is necessary to 
define the notion of partial recursive function. We limit ourselves to 
give an intuitive idea which is very close to the formal 
definition. We can consider a partial recursive function as a computer 
$C$ which takes a program $P$ (namely a binary string) as an input, 
performs some computations and gives a string $s=C(P)$, written in the 
given alphabet $\mathcal{A}$, as an output. The $AIC$ of a string $s$ 
is defined as the shortest binary program $P$ which gives $s$ as its 
output, namely 
\begin{equation*} 
{AIC}(s,C)=\min \{|P|:C(P)=s\} 
\end{equation*} 
 
We require that our computer is a universal computing machine. Roughly 
speaking, a computing machine is called \emph{universal} if it can 
simulate any other machine. In particular every real computer is a 
universal computing machine, provided that we assume that it has 
virtually infinite memory. For a precise definition see 
e.g. \cite{livi} or \cite{cha}. We have the following theorem due to 
Kolmogorov (\cite{kolmogorov},\cite{livi}). 
 
\begin{theorem} 
If $C$ and $C^{\prime }$ are universal computing machines then  
\begin{equation*} 
\left| {AIC}(s,C)-{AIC}(s,C^{\prime })\right| \leq K\left( C,C^{\prime 
}\right) 
\end{equation*} 
where $K\left( C,C^{\prime }\right) $ is a constant which depends only 
on $C$ and $C^{\prime }$ but not on $s$. 
\end{theorem} 
 
This theorem implies that the information content ${AIC}$ of $s$ with 
respect to $C$ depends only on $s$ up to a fixed constant, then its 
asymptotic behavior does not depend on the choice of $C$. For this 
reason from now on we will write ${AIC}(s)$ instead of ${AIC}(s,C).$ 
The shortest program which gives a string as its output is a sort of 
encoding of the string. The information which is necessary to 
reconstruct the string is contained in the program. 
 
We have the following result (for a proof see for example 
\cite{galDCDS} Lemma 6) : 
 
\begin{theorem} 
\label{222} Let  
\begin{equation*} 
Z_{C}:\mathcal{A}^{\ast }\rightarrow \left\{ 0,1\right\} ^{\ast } 
\end{equation*} 
be the function which associates to a string $s$ the shortest program 
whose output is $s$ itself\footnote{It 
two programs of the same length produce the same string, we choose the 
program which comes first in lexicographic order.} 
(namely, $AIC(s)=I_{Z_{C}}(s)$). If $Z$ is any reversible coding, 
there exists a constant $M$ which depends only on $C$ such that 
\begin{equation} 
\left| Z_{C}\left( s\right) \right| \leq \left| Z\left( s\right) \right| +M 
\label{1} 
\end{equation} 
\end{theorem} 
 
The inequality (\ref{1}) says that $Z_{C}\;$in some sense is optimal. 
Unfortunately this coding procedure cannot be performed by any 
algorithm (Chaitin Theorem)\footnote{Actually, the Chaitin theorem 
states a weaker statement: a procedure (computer program) which states 
that a string $\sigma $ of length $n$ can be produced by a program 
shorter than $n,$ must be longer than $n.$}. This is a very deep 
statement and, in some sense, it is equivalent to the Turing halting 
problem or to the G\"odel incompleteness theorem. Then the Algorithmic 
Information Content is not computable by any algorithm. 
 
This fact has very deep consequences for our discussion as we will see
later. For the moment we can say that the $AIC$ cannot be used as a
reasonable physical quantity since it cannot be measured, however it
is very useful in proving general theorems.
 
\section{Information sources} 
 
\subsection{Infinite strings and complexity} 
 
A symbolic dynamical system is given by 
$(\Omega,\mathcal{C},\mu,\sigma)$. The space $\Omega$ is the space 
$\mathcal{A}^{\mathbb{N}}$ of the infinite sequences 
$\omega=(\omega_{i})_{i\in\mathbb{N}}$ of symbols in $\mathcal{A}$. 
$\mathcal{C}$ is the $\sigma$-algebra generated by the 
cylinders\footnote{We remark that $\mathcal{C}$ corresponds to the 
Borel $\sigma$-algebra when $\Omega$ is equipped with the product 
topology, that is the topology induced by the metric $d(\omega, 
\overline{\omega})=\sum_{i\in \mathbb{N}} \ \frac{ 
\delta(\omega_i,\overline{\omega}_i)}{2^i}$, where 
$\delta(\cdot,\cdot)$ is the Kronecker delta.}, 
\begin{equation*} 
C(\omega ^{(k,n)})=\{\overline{\omega }\in \Omega :\overline{ \omega } 
_{i}=\omega _{i}\ for\ k\leq i\leq n-1\}, 
\end{equation*} 
where $\omega ^{(k,n)}=(\omega _{i})_{k\leq i\leq n-1}=(\omega 
_{k},\omega _{k+1},\dots ,\omega _{n-1})$, the map $\sigma$ is the 
shift map 
\begin{equation*} 
\sigma((\omega_{i})_{i\in\mathbb{N}})=(\omega_{i+1})_{i\in\mathbb{N}}  
\end{equation*} 
and $\mu$ is a $\sigma$-invariant probability measure on $\Omega$. A 
symbolic dynamical system can be also viewed as an information 
source. For the purposes of this work the two notions can be 
considered equivalent. 
 
We give now different measures of complexity for infinite strings 
generated by the symbolic dynamical system according to Definition 
\ref{infmeas}, using the different information measures defined 
above. However each definition of complexity of an infinite string 
$\omega$ can be thought of as a measure of the average quantity of 
information which is contained in a single digit of $\omega$. 
 
\begin{definition}[Complexity of infinite strings] 
If $\omega\in\Omega$, $Z:\mathcal{A}^*\rightarrow \{0,1\}^*$ is a 
reversible universal coding procedure we define the {\it computable 
complexity} of $\omega$ with respect to $Z$ as 
\begin{equation*} 
K_{Z}(\omega)=\mathrel{\mathop{limsup}\limits_{n\rightarrow\infty}} 
K_Z(\omega ^n) \ , 
\end{equation*} 
where $\omega^n= \omega^{(0,n)}$. In the same way, using the $AIC$, we 
define 
\begin{equation*} 
K(\omega )= \mathrel{\mathop{limsup}\limits_{ n\rightarrow \infty }} \frac {{ 
{AIC}}(\omega ^n)}n\ .  
\end{equation*} 
\end{definition} 
 
We also define the quantity $\hat{H}(\omega)$. If $\omega$ is an 
infinite string, $\hat{H}(\omega)$ is a sort of Shannon entropy of the 
single string. 
 
\begin{definition} 
By the definition of empirical entropy of finite strings we define:  
\begin{equation*} 
\hat{H}_{l}(\omega)=\mathrel{\mathop{limsup}\limits_{n\rightarrow\infty} 
}\hat{H}_{l}(\omega^{n})  
\end{equation*} 
and  
\begin{equation*} 
\hat{H}(\omega)=\mathrel{\mathop{lim}\limits_{l\rightarrow\infty}} \hat{H} 
_{l}(\omega).  
\end{equation*} 
\end{definition} 
 
The existence of this limit is proved in \cite{lz78}. The following 
proposition is a direct consequence of ergodicity (for the proof see 
again \cite{lz78}). 
 
\begin{proposition} 
\label{6} If $(\Omega,\mu,\sigma)$ is ergodic then $\hat{H}(\omega)=h_{\mu} 
(\sigma)$ (where $h_{\mu}$ is the Kol\-mo\-go\-rov-Sinai entropy of 
$\sigma$) for $\mu $-almost each $\omega$. 
\end{proposition} 
 
Moreover from the definition of optimality it directly follows that: 
 
\begin{remark} 
\label{76} If $Z$ is optimal then for each $\omega$ and for all $l$  
\begin{equation*} 
K_{Z}(\omega)\leq\hat{H}_{l}(\omega)\ ,  
\end{equation*} 
so that  
\begin{equation*} 
K_{Z}(\omega)\leq\hat{H}(\omega)\ .  
\end{equation*} 
\end{remark}

\begin{remark} 
\label{77} As it is intuitive, the compression ratio of $Z$ cannot be less 
than the average information per digit as it is measured by the algorithmic 
information content (Theorem \ref{222}), thus for all $\omega $, we have  
\begin{equation*} 
K_{Z}(\omega)\geq K(\omega)\ . 
\end{equation*} 
\end{remark} 
 
This remark and the following Lemma are useful for the proof of the 
next theorem 
 
\begin{lemma}[Brudno \cite{brud}]\label{brud1} If $\mu $ is ergodic then  $K(\omega)=h_\mu(\sigma)$ for almost each $\omega$. 
\end{lemma}

Then we have the following 
 
\begin{theorem} 
\label{11} If $(\Omega,\mu,\sigma)$ is a symbolic dynamical system, $Z$ is 
optimal and $\mu$ is ergodic, then for $\mu$-almost each $\omega$  
\begin{equation*} 
K_{Z}(\omega)=\hat{H}(\omega)=K(\omega)=h_{\mu}(\sigma)\ ,  
\end{equation*} 
in particular optimality implies that the algorithm is an ideal coding 
and $I_Z$ and ${AIC}$ are information measures (see Definition 
\ref{infmeas}). 
\end{theorem} 
 
\begin{proof} 
$\hat {H}(\omega )=K(\omega )=h_{\mu}(\sigma)$ for almost each 
$\omega$ using Proposition \ref{6} and Brudno's Lemma above. Moreover 
we have that $K_{Z}(\omega)\geq K(\omega)$ (Remark \ref{77}) and then 
$K_{Z}(\omega)\geq h_{\mu}(\sigma)$ for $\mu $-almost each $\omega$. 
 
On the other hand, $K_{Z}(\omega)\leq\hat{H}(\omega)$ (Remark 
\ref{76}) and then $K_{Z}(\omega)=h_{\mu}(\sigma )$ for almost each 
$\omega$. 
\end{proof} 
 
This  theorem shows that all the various information measures we have
defined in section 2 agrees when we study the long time asymptotical behavior
of the information necessary to describe a generic orbit of a positive entropy source.

If the measure $\mu $ is not ergodic we can replace the a.e. above result 
with an average result: the average complexity is equal to the entropy.

\begin{theorem} 
\label{teomedia} Let $(\Omega,\mathcal{C},\sigma)$ be a symbolic dynamical 
system, with a $\sigma$-inva\-riant probability measure $\mu$. Then if $Z$ 
is optimal,  
\begin{equation*} 
h_{\mu} (\sigma)=\int_{\Omega} K_Z(\omega) d\mu =\int_{\Omega} \hat H 
(\omega) d\mu =\int_{\Omega} K(\omega) d\mu\ . 
\end{equation*} 
\end{theorem} 
 
\begin{proof} 
First of all, we show that all the quantities to be integrated are 
actually measurable. We show how to prove measurability for 
$K(\omega)$. The argument applies unchanged to the others (one more 
limit has to be considered for $\hat H$). This argument is due to 
Brudno (\cite{brud}). 
 
For any $t\in \mathbb{R}$, let $T=\{ \omega \in \Omega \ / \ K(\omega) 
< t \} $. The set $T$ can be written as 
\begin{equation*} 
T = \bigcup_{k=1}^\infty \ \bigcup_{N=1}^\infty \ \bigcap_{n>N} \ \{ \omega 
\ / \ AIC(\omega^n) < n(t-1/k) \}, 
\end{equation*} 
and since all the sets in curly brackets are finite union of 
cylinders, measurability of the set $T$ and of $K(\omega)$ follows from 
classical theorems of measure theory. 
 
To obtain the thesis of the theorem, we use the \textit{ergodic 
decomposition theorem} and its application to Kolmogorov-Sinai entropy 
$ h_\mu (\sigma)$ (see Katok-Hasselblatt, chapter 4, \cite{kh}). Let $ 
(\Omega_j,\mathcal{C}_j,\mu_j)_{j\in J}$ be an ergodic decomposition 
of $ (\Omega,\mathcal{C},\sigma)$, that is $\Omega_j$ are invariant 
subsets of $ \Omega$, $\mu_j$ are ergodic measures with support on 
$\Omega_j$, and $J$ is a Lebesgue space with probability measure 
$P$. Then we have that 
\begin{equation*} 
\int_{\Omega} K(\omega) d\mu = \int_J \left( \int_{\Omega_j} \ K(\omega) \ d 
\mu_j \right) \ dP = \int_J \ h_{\mu_j} (\sigma) \ dP = h_\mu (\sigma). 
\end{equation*} 
The first and last equalities come from the ergodic decomposition theorem, 
and the second one from Theorem \ref{11}. The same argument applies to $ 
K_Z(\omega)$, $\hat H(\omega)$. 
\end{proof} 
 
\section{Dynamical systems} 
 
In this Section we apply the features of coding algorithms  
and the results of the previous section to define a 
notion of complexity for orbits of dynamical systems and prove some 
relations with the Kolmogorov-Sinai entropy. 
 
The relations we can prove will be useful as a theoretical 
support for the interpretation of the experimental and numerical 
results.  The results which we will explain in this section are 
meaningful in the positive entropy case. The null entropy cases are 
harder to deal with, and we present some results in the next section. 
 
\subsection{Dynamical systems and partitions} 
 
\label{dynsyst} 
 
Now we consider a dynamical system $(X,\mu ,T)$, where $X$ is a 
compact metric space, $T$ is a continuous map $T:X\rightarrow X$ and 
$\mu $ is a Borel probability measure on $X$ invariant for $T$. If 
$\alpha =\{A_{1},\dots ,A_{n}\}$ is a measurable partition of $X$ (a 
partition of $X$ where the sets are measurable) then we can associate 
to $(X,\mu ,T)$ a symbolic dynamical system $(\Omega _{\alpha },\mu 
_{\alpha })$ ( called a symbolic model of $(X,T)$). By this 
association many results about symbolic dynamical systems will be 
translated to dynamical systems over metric spaces where the choice of 
a partition has been made. 
 
The set $\Omega _{\alpha }$ is a subset of $\{1,\dots 
,n\}^{\mathbb{N}}$ (the space of infinite strings made of symbols from 
the alphabet $\{1,\dots ,n\}$). To a point $x\in X$ it is associated a 
string $\omega =(\omega _{i})_{i\in \mathbb{N}}=\varphi _{\alpha }(x)$ 
defined as 
\begin{equation*} 
\varphi _{\alpha }(x)=\omega \iff \forall j\in \mathbb{N}, \  
\ T^{j}(x)\in A_{\omega _{j}}. 
\end{equation*} 
Since ${\alpha }$ is a partition the set $\varphi_{\alpha}(x)$ will 
contain only one element and defines a function associating an 
infinite string to a point $x\in X$. The measure $\mu $ on $X$ induces 
a measure $\mu _{\alpha }$ on the associated symbolic dynamical 
system. The measure is first defined on the cylinders\footnote{ We 
recall that $\omega ^{(k,n)}=(\omega _{i})_{k\leq i\leq n}=(\omega 
_{k},\omega _{k+1},\dots ,\omega _{n})$.} 
\begin{equation*} 
C(\omega ^{(k,n)})=\{\overline{\omega }\in \Omega _{\alpha }:\overline{ 
\omega }_{i}=\omega _{i}\ for\ k\leq i\leq n-1\} 
\end{equation*} 
by  
\begin{equation*} 
\mu _{\alpha }(C(\omega ^{(k,n)}))=\mu (\cap _{k}^{n-1}T^{-i}(A_{\omega 
_{i}})) 
\end{equation*} 
and then extended by the classical Kolmogorov theorem about product 
measures to a measure $\mu_{\alpha }$ on $\Omega_{\alpha }$. Moreover 
if $(X,\mu ,T)$ is ergodic then $(\Omega_{\alpha },\mu_{\alpha 
},\sigma)$ is ergodic and $h_{\mu }(T,\alpha )$ (the Kolmogorov-Sinai 
entropy relative to the partition $\alpha$) on $X$ equals 
$h_{\mu_{\alpha }}(\sigma)$ on $\Omega_{\alpha }$ (see also 
\cite{brud}). 
 
We now define the complexity of an orbit with respect to a 
partition. The above considerations will allow us to apply the results 
on symbolic dynamical systems to general dynamical systems with a 
partition. 
 
\begin{definition} 
Let $\omega=\varphi_\alpha (x)$ for a given partition $\alpha$. We 
define the {\it complexity of the orbit of a point} $x\in X$, with 
respect to the partition $\alpha $, as 
 
\begin{equation*} 
AIC(x,\alpha ,n)=AIC(\omega ^{n}) 
\end{equation*} 
\begin{equation*} 
K(x,\alpha )=K(\omega ), 
\end{equation*} 
where the information is measured by the AIC. We also define 
 
\begin{equation*} 
I_{Z}(x,\alpha ,n)=I_{Z}(\omega ^{n}) 
\end{equation*} 
\begin{equation*} 
K_{Z}(x,\alpha )=K_{Z}(\omega ), 
\end{equation*} 
where the information is measured by $Z$. Also the definition of 
empirical entropy can be extended for $(x,\alpha )$, defining 
 
\begin{equation*} 
\hat{H}(x,\alpha)=\hat{H}(\omega )\ .  
\end{equation*} 
\end{definition} 
 
\begin{theorem} 
\label{14} If $Z$ is an optimal coding, and $(X,\mu ,T)$ is an ergodic 
dynamical system and $\alpha $ is a measurable partition of $X$, then for $ 
\mu$-almost all $x$  
\begin{equation*} 
K_{Z}(x,\alpha )=h_{\mu }(T,\alpha ) 
\end{equation*} 
where $h_{\mu }(T,\alpha )$ is the Kolmogorov entropy of $(X,\mu ,T)$ with 
respect to the measurable partition $\alpha $. 
\end{theorem} 
 
The proof of the above Theorem follows easily from the following 
lemmas. 
 
\begin{lemma} \label{ciao} 
If $(X,T,\mu)$ is ergodic, then for almost each point $x \in X$, 
$K(x,\alpha)=h_{\mu}(T,\alpha)$. 
\end{lemma} 
 
This Lemma was already proved by Brudno. See \cite{brud} Lemma 2.6 
page 137. 
 
\begin{lemma} 
\label{hat} If $(X,T,\mu)$ is ergodic, for almost each point $x \in X$  
\begin{equation*} 
\hat{H}(x,\alpha )=h_\mu (T,\alpha). 
\end{equation*} 
\end{lemma} 
 
\begin{proof} 
In the associated symbolic system 
$h_{\mu}(T,\alpha)=h_{\mu_\alpha}(\sigma )$. Moreover, for almost each 
$\omega \in \Omega _\alpha $, it holds $\hat {H}(x,\alpha 
)=\hat{H}(\omega )=h_{\mu _\alpha}(\sigma)$ where $x= { 
\varphi_{\alpha}}^{-1}(\omega) $ (Prop. \ref{6}). If we consider 
$Q_{\Omega_{\alpha}}:=\{\omega\in\Omega_{\alpha}:\hat{H} 
(\omega)=h_{\mu_{ \alpha}}(\sigma)\}$ and $Q:=\varphi_{\alpha 
}^{-1}(Q_{\Omega_{\alpha}})$ we have 
\begin{equation*} 
\forall x\in Q\quad \hat{H}(x,\alpha)=\hat{H}(\varphi_{\alpha}(x))=h_{\mu_{ 
\alpha} }(\sigma)=h_{\mu}(T,\alpha)\ .  
\end{equation*} 
According to the way in which the measure $\mu_{\alpha}$ is constructed we 
have $\mu(Q)=\mu_{\alpha}(Q_{\Omega_{\alpha}})=1$. 
\end{proof} 
 
{\em Proof of Theorem \ref{14}.} The proof of Theorem \ref{14} follows as before from the remark that 
$K(x,\alpha)\leq K_Z(x,\alpha )\leq \hat{H}(x,\alpha)$ (Remarks 
\ref{76} and \ref{77}) and Lemmata \ref{ciao} and \ref{hat}.$\Box$
 
As before we show the corresponding result in the non ergodic case 
 
\begin{theorem} 
\label{cne} If for the dynamical system $(X,T,\mu)$ the measure $\mu$  
is only $T$-invariant, then, if $Z$ is an optimal compression 
algorithm, for any measurable partition $\alpha$ it holds 
\begin{equation*} 
h_{\mu} (T,\alpha)=\int_{X} K_Z(x,\alpha) \ d\mu =\int_{X} \hat H 
(x,\alpha) \ d\mu =\int_{X} K(x,\alpha) \ d\mu\ . 
\end{equation*} 
\end{theorem} 
 
\begin{proof} 
The proof follows that of Theorem \ref{teomedia}, using the definition 
of the complexity of infinite orbits of a dynamical system through the 
complexity of the associated infinite symbolic orbit, and previous 
lemmata. The measurability of the partition $\alpha$ is essential to 
obtain the measurability of the function $\varphi_\alpha:X\to 
\Omega_\alpha$.  
\end{proof}

\begin{corollary} Under the assumption of the previous Theorem \ref{cne}, if
  moreover $\alpha$ is a generating partition
\begin{equation*} 
h_{\mu} (T)=\int_{X} K_Z(x,\alpha) \ d\mu =\int_{X} \hat H 
(x,\alpha) \ d\mu =\int_{X} K(x,\alpha) \ d\mu\ . 
\end{equation*} 
\end{corollary} 

\textbf{Remarks.}
This  theorem shows that all the various information measures we have
defined in section 2 agrees when we study the long time asymptotical behavior
of the information necessary to describe a generic orbit of a positive entropy
system.  Theorem \ref{cne} shows that if a system has an 
invariant measure, its entropy with respect to a given partition can 
be found by averaging the complexity of its orbits over the invariant 
measure. Then, the entropy may be alternatively defined as the average 
orbit complexity.  However if we fix a single point, its orbit 
complexity is not yet well defined because it depends on the choice of 
a partition. It is not possible to get rid of this dependence by 
taking the supremum over all partitions (as in the construction of 
Kolmogorov-Sinai entropy), because this supremum goes to infinity for 
each orbit that is not eventually periodic (see \cite{brud} Assertion 
2.8). 
 
We sketch how this difficulty may be overcome in two ways: 
 
1) by considering open covers instead of partitions as in \cite{brud}, 
\cite{gal2} and in \cite{galDCDS}. We recall that since the sets in an 
open cover can have non empty intersection, a step of the orbit of $x$ 
can be contained at the same time in more than one open set of the 
cover. This implies that an orbit may have an infinite family of 
possible symbolic codings, among which we choose the ``simplest one''. 
Then we can define the complexity of the orbit of a point as the 
supremum of the complexities obtained with respect to all possible 
open covers. This definition has the very nice property to be 
invariant up to topological equivalence of dynamical systems. This 
definition of orbit complexity equals the entropy for almost each 
point of a compact ergodic system. 
 
2) by considering only a particular class of partitions and define the 
orbit complexity of a point as the supremum of the orbit complexity 
over that class.  This can be easily done if the space is $R^n$ by 
considering partitions generated by intersections of half spaces with 
rational coordinates (polyedric partitions). By the following Lemma 
\ref{1122} it easily follows that the corresponding notion of orbit 
complexity equals the entropy for almost each point of a compact 
ergodic system. 
 
\vskip0.5cm Let $\beta_{i}$ be a family of measurable partitions such 
that $ \mathrel{\mathop{\lim}\limits_{ 
i\rightarrow\infty}}\,diam(\beta_{i})=0$. If we consider $ 
\mathrel{\mathop{\lim\sup 
}\limits_{i\rightarrow\infty}}K_{Z}(x,\beta_{i})$ we have the 
following 
 
\begin{lemma} 
\label{1122} If $(X,\mu,T)$ is compact and ergodic, $Z$ is optimal, then for  
$\mu$-almost all points $x\in X$, $ \mathrel{\mathop{\limsup 
}\limits_{i\rightarrow\infty}}K_{Z}(x,\beta_{i})= 
\mathrel{\mathop{\lim\sup } 
\limits_{i\rightarrow\infty}}K(x,\beta_{i})=h_{\mu}(T)$. 
\end{lemma} 
 
\begin{proof} 
The points for which $K_{Z}(x,\beta_{i})\neq h_{\mu}(T,\beta _{i})$ 
are a set of null measure for each $i$ (Theorem \ref{14}). When 
excluding all these points, we exclude (for each $i$) a zero-measure 
set. For all the other points we have 
$K_{Z}(x,\beta_{i})=h_{\mu}(T,\beta_{i})$ and then $ 
\mathrel{\mathop{\lim\sup 
}\limits_{i\rightarrow\infty}}K_{Z}(x,\beta_{i})= 
\mathrel{\mathop{\lim\sup }\limits_{i\rightarrow\infty 
}}h_{\mu}(T,\beta_{i})$. Since $X$ is compact and the diameter of the 
partitions $\beta_{i}$ tends to 0, we have that $ 
\mathrel{\mathop{\lim\sup 
}\limits_{i\rightarrow\infty}}h_{\mu}(T,\beta_{i})=h_{\mu}(T)$ (see 
e.g.  \cite{kh} page 170). The same arguments holds for 
$K(x,\beta_{i})$, and the statement is proved. 
\end{proof} 

The previous lemma makes possible the following definition.  If
$(X,\mu,T)$ is compact and ergodic and $Z$ is optimal, then for
$\mu$-almost all points $x\in X$ and for countable families of
measurable partitions $\{\beta _i\}_{i\in\mathbb{N}}$ such that $
\mathrel{\mathop{\lim}\limits_{
i\rightarrow\infty}}\,diam(\beta_{i})=0$, the {\bf complexity of the
orbit of a point $x\in X$} is $$K_Z(x)=\mathrel{\mathop{\lim\sup
}\limits_{i\rightarrow\infty}}K_{Z}(x,\beta_{i})\ .$$

\section{Weakly chaotic dynamical systems\label{WCS}} 
 
A weakly chaotic dynamical system is a system whose all physically
relevant invariant measures have null Kolmogorov-Sinai entropy, but it
has a not ordered dynamics. Thus, the complexity defined in
Def. \ref{infmeas} always gives a null value and it is not a good
observable to characterize these systems.
 
The first thing to do to have a meaningful observable would be to look 
directly at the asymptotic behavior of the information necessary to 
describe the orbit of a point\footnote{We recall that we do this with  
respect to a fixed  partition (see the remarks of the 
previous section)}. One of the main tools in the proof of Brudno's 
main theorem, which states the equality between the complexity for 
almost any initial condition and the Kolmogorov-Sinai entropy, is the 
Birkhoff ergodic theorem, which gives a relation between spatial and 
temporal averages for measurable functions defined on the state 
space. Then our pointwise approach to the asymptotic behavior of the 
complexity corresponds to the pointwise results of the ergodic 
theorem.  
 
From this point of view, the pointwise approach in weakly chaotic
dynamical systems should be based on general ergodic theorems, in
which the temporal average should be done with non linear
weights. This is, to our knowledge, a very delicate point in ergodic
theory, and actually there are some negative results, for example in
case of dynamical systems defined on a space $X$ with an invariant
measure $\mu$ such that $\mu(X)=+\infty$. Let $(X,T,\mu)$ be such a
dynamical system; it is impossible to define a sequence $\{ a(n) \}$
of integer numbers, monotonically converging to infinity and with
$\frac{a(n)}{n} \to 0$ as $n\to \infty$, such that for all functions
$f\in L^1(X,\mu)$
$$\lim_{n\rightarrow
+\infty}\frac{1}{a(n)}\sum_{i=0}^{n-1}f(T^i(x))=C$$ for almost any
$x\in X$, where $C$ is a positive finite constant
(\cite{Aaronson}). This result is applicable, for example, to the
family of Manneville maps with parameter $z\ge 2$ (see Sections
\ref{man77} and \ref{numMan} for the description of the maps and
references \cite{GW}, \cite{gal3},\cite{claudio}) where the physically
relevant invariant measure is infinite. Hence this is an indication
that a pointwise approach for the complexity of the orbits of the
Manneville maps could not give a consistent result.
 
We remark that we are just looking at the behavior of ergodic averages 
of a single function, so the generality of the ergodic theorems could 
be too much for our aims. Nevertheless using the results of 
 \cite{isola} and \cite{colfer} for the Manneville map with $z=2$, we 
expect that for almost any point $x\in X$ it is impossible to find a 
sequence $a(n)$ of integer numbers, converging to infinity and with 
$\frac{a(n)}{n} \to 0$ as $n\to \infty$, such that the limit 
$$ 
\lim\limits_{n \to \infty} \frac{AIC(\omega^n)}{a(n)} 
$$ 
exists and it is strictly positive, where $\omega^n$ denotes the first 
$n$ digits of the symbolic string associated to the point $x$ using a 
fixed partition. Moreover we expect the superior limit to be infinity 
and the inferior limit to be zero for almost any initial condition, 
when the two limits are not both either zero or infinity. We believe 
that this result can be extended to the cases $z>2$. 
 
Hence, for this reason, we will suggest a slight modification of 
Definition \ref{infmeas} and we will show that in the case of the 
Manneville-type maps this new index gives a classification for the maps 
of the family. 
 
First, we will sketch the landscape in the case of general symbolic dynamical 
systems. 
 
\subsection{Symbolic dynamical systems} 
 
The following definitions are inspired by the example of the 
Manneville maps with $z>2$, for which there is not a physically 
relevant invariant probability measure, but for which there are 
results about the Lebesgue measure which is physically relevant but not invariant.
 
Hence in the following we will consider dynamical systems with a not
necessarily invariant reference probability measure $\mu$. Let
$(\Omega,\sigma)$ be a dynamical system and assume that there is a
physically relevant measure $\mu$ which is not necessarily
invariant. For instance, if the space $X$ is the unit interval
$[0,1]$, then we will consider $\mu$ to be the Lebesgue measure on
$[0,1]$. We consider the following index:
 
\begin{definition}[q-entropy] 
\label{accaqAICZ} Let $I:\mathcal{A}^*\rightarrow\mathbb{N}$ be an 
information measure. Let $q$ be a positive real number. We call 
\textit{q-entropy} 
\begin{equation} \label{accaq} 
h^q (\Omega) = \limsup\limits_{n\rightarrow +\infty} \int_{\Omega}\frac{ 
I(\omega^n)}{n^q}\ d\mu\ . 
\end{equation} 
If $I={AIC}$, then we denote $h^q (\Omega)$ with $h^q_{AIC} 
(\Omega)$. If $ I=I_Z$, with $Z$ a recursive coding procedure, then we 
denote $h^q (\Omega)$ with $h^q_Z (\Omega)$. 
\end{definition} 
 
\begin{theorem} 
\label{teoaccaq} For all recursive coding procedures $Z$ and all $q>0$, 
we have 
\begin{equation*} 
h^q_Z(\Omega) \geq h^q_{AIC}(\Omega)\ . 
\end{equation*} 
\end{theorem} 
 
\begin{proof} 
From inequality (\ref{1}), we have that there exists a constant $M$ not 
depending on $Z$ such that  
\begin{equation*} 
\frac{{AIC}(\omega^n)}{n^q} \leq \frac{I_Z(\omega^n)}{n^q} + 
\frac{M}{n^q} \ . 
\end{equation*} 
From this the theorem easily follows. 
\end{proof} 
 
\begin{definition}[Chaos index] 
\label{chaind} We call {\it chaos index} of the symbolic dynamical system  
$(\Omega, \mathcal{C},\mu,\sigma)$ the number $q(\Omega)=\inf \{ p>0 \ 
| \ h^p(\Omega)=0 \} \in [0,1]$. The indexes $q_{_{AIC}}$ and $q_{_Z}$ 
are defined as above. 
\end{definition} 
 
\begin{corollary} 
For all recursive coding procedures $Z$,   
\begin{equation*} 
q_{_Z}(\Omega) \geq q_{_{AIC}}(\Omega)\ . 
\end{equation*} 
\end{corollary} 
 
\subsection{General dynamical systems} 
 
Let $(X,T)$ be a dynamical system amnd let $\mu$ be a reference probability
measure as above, which is not supposed to be invariant. Let $
(\Omega_\alpha,\mu_\alpha)$ be a symbolic model of $(X,T)$, relative
to the partition $\alpha$ of the space $X$ and $\varphi_\alpha (x)$
the symbolic string associated to any point $x\in X$.
 
\begin{definition}[q-entropy relative to a partition] 
\label{accaqpart} As above let $I:\mathcal{A}^*\rightarrow\mathbb{N}$ be an 
information measure, either ${AIC}$ or $I_Z$. Let $\alpha$ be a partition 
of $X$ and $I(x,\alpha,n)=I(\omega^n)$, where $\omega = \varphi_\alpha (x)$. 
Then we call {\it q-entropy relative to the partition $\alpha$}  
\begin{equation}  \label{accapart} 
h^q (X,\alpha) = \limsup\limits_{n\rightarrow +\infty} \int_{\Omega_\alpha} 
\frac{I(x,\alpha,n)}{n^q}\ d\mu_\alpha \ . 
\end{equation} 
\end{definition} 
 
In the example of the family of Manneville maps, that is the simplest 
model of intermittent weak chaos, the average with respect to the 
Lebesgue measure of the information plays a crucial role in the 
classification of the maps in the family. Following this example, we 
believe that the following index is a particularly meaningful 
indicator in the study of intermittent weakly chaotic dynamical 
systems. 
 
\begin{definition}[Intermittent Chaos index] 
\label{indchaos}   
We call {\it intermittent chaos index} of the dynamical system 
$(X,T,\mu)$ with respect to a partition $\alpha $ the number 
\begin{equation*} 
q(X,T,\alpha )=\inf \{ p>0 \ | \ h^p(X,\alpha)=0 \} \in [0,1] \ . 
\end{equation*} 
The indexes $q_{_{AIC}}$ and $q_{_Z}$ are defined as above. 
\end{definition} 
 
\begin{corollary} 
For all  compression algorithms $Z$,   
\begin{equation*} 
q_{_Z}(X,T,\alpha ) \geq q_{_{AIC}}(X,T,\alpha )\ . 
\end{equation*} 
\end{corollary} 
 
In the next section we will apply these definitions to the family 
Manneville maps (for the definition see the next section) choosing the 
$LZ77$ compression algorithm (\cite{lz77}, \cite{lz78}) and a 
generating partition $\alpha$. As a result it holds that in the weakly 
chaotic case (for $z> 2$), 
\begin{equation*} 
q_{_{LZ77}}(X,T,\alpha )\ =\ q_{_{AIC}}(X,T,\alpha )\ =\ \frac {1}{z-1}\ . 
\end{equation*}

\section{Compression algorithms}\label{comal} 
 
\subsection{The algorithm $LZ77$} 
The Ziv-Lempel compression scheme LZ77 with infinite window 
(\cite{lz77}) is the one from which almost all practical adaptive 
dictionary encoders derived (\cite{bell}). A dictionary of an input 
string is the set of words (i.e. group of consecutive symbols) in 
which the algorithm parses the input string. 
 
The essence is that phrases (i.e. sets of consecutive words in the string 
to be encoded) are replaced with a pointer to where they have occurred 
earlier in the input string. Novel words and phrases can also be constructed 
from parts of earlier words. 
 
In the $LZ77$ compression algorithm, the new word is defined as a pair 
({\it pointer},{\it symbol}). The pointer is referred to a phrase 
contained in the part of the input string which precedes the current 
position of the front end. As an example, let the alphabet 
$\mathcal{A}$ be the set $\{a_{1},\dots ,a_{r}\}$ and consider an 
input string $\omega \in 
\mathcal{A}^{\ast }$. As usual, $\omega ^{n}=(\omega _{1}\cdots \omega 
_{n})$ is the substring of $\omega $ of length $n$ and containing its 
first $n$ symbols. \newline 
\indent Consider some step $h$ of the coding procedure. 
 
Suppose the first $p$ symbols $(\omega _{1}\cdots \omega _{p})$ have 
already been encoded. The dictionary now contains $h$ words 
$\{e_1,\dots,e_h\}$. Thus, the current position of the front end is 
the $(p+1)$$^{th}$ site in the input string and the next word in the 
dictionary will be labelled as the $(h+1)$$^{th}$ word $e_{h+1}$. 
 
The algorithm selects this new word as the longest word which can be 
obtained by adding a single character $\tilde{a}$ chosen in the 
alphabet $\mathcal{A}$ to a phrase $\rho$ contained in the substring 
$(\omega_1\cdots\omega_{p-2})$. Hence, the word $e_{h+1}$ 
has as a prefix the phrase $\rho$ followed by $\tilde{a}$ as an ending 
symbol ($e_{h+1}=\rho\ \tilde{a}$). 
 
Once the new word $e_{h+1}$ has been found, the algorithm outputs a 
binary encoding of the triplet $(s_{h+1},l_{h+1},\tilde{a})$ where 
$s_{h+1}$ is the starting position of the prefix $\rho$ of the new 
word in the string $(\omega _{1}\cdots \omega _{p})$, $l_{h+1}$ is the 
length of the new word $e_{h+1}$ and the symbol $\tilde{a}$ from the 
alphabet is the last character of $e_{h+1}$. 
 
The following example shows how the algorithm $LZ77$ encodes the input 
stream  
\begin{equation*} 
\omega\ =\ (aababbbbaababba\dots)\ .  
\end{equation*} 
\indent Let $\mathcal{A}=\{a,b\}$ be the source alphabet. 
 
The output is the binary encoding of the following triplets. The first 
column is the dictionary index number of the codeword whose triplet is 
showed in the same line, second column. For an easier reading, we add a 
third column which shows each encoded word in the original stream $s$, but 
we remark that it is not contained in the output file:  
\begin{equation*} 
\begin{array}{lll} 
1 & (1,1,\ ^{\prime}a\ ^{\prime}\ ) & [a] \\  
2 & (1,2,\ ^{\prime}b\ ^{\prime}\ ) & [ab] \\  
3 & (2,3,\ ^{\prime}b\ ^{\prime}\ ) & [abb] \\  
4 & (5,3,\ ^{\prime}a\ ^{\prime}\ ) & [bba] \\  
5 & (2,5,\ ^{\prime}a\ ^{\prime}\ ) & [ababba] 
\end{array} 
\end{equation*} 
and so on. 
 
Now we will recall some results from \cite{manzini}, concerning the 
optimality of the $LZ77$ algorithm. 
 
\begin{lemma} 
\label{lemmat} Let $t$ denote the number of words in which the algorithm  
$LZ77$ parses the string $\omega ^{n}$. Set $m=n-N$ where 
$N=\max_{i=1,\dots ,r}n_{i}$, where $n_i$ is the number of 
occurrences of the symbol $a_i\in {\mathcal A}$. Then $t\leq 2(m+1)$. 
\end{lemma} 
 
\begin{theorem}\label{38} 
\label{teomanz} Let $t$ denote the number of words in which the algorithm  
$LZ77$ parses the string $\omega ^{n}$. 
 
\begin{description} 
\item (i) For all $k\geq 1$ it holds 
\begin{equation} 
\begin{array}{ll} 
I_{LZ77}(\omega ^{n})=|LZ77(\omega ^{n})| & \leq \ n\ \hat{H}_{k}(\omega 
^{n})+3\ t\log _{2}\Big(\frac{n}{t}\Big)+ \\[15pt]  
& +\ O\Big((k-1)\ t+t\log _{2}\log _{2}\Big(\frac{n}{t}\Big)\Big)\ . 
\end{array} 
\label{lowerorder} 
\end{equation} 
 
\item (ii) The algorithm $LZ77$ is optimal and for all $n\geq 1$, for 
all $\omega^{n}\in \mathcal{A}^{n}$ and for all $k\geq 1$ it holds 
\begin{equation} 
\frac{I_{LZ77}(\omega ^{n})}{n}\leq \ \hat{H}_{k}(\omega ^{n})+O\Bigg(\frac{ 
\log _{2}\log _{2}(n)}{\log _{2}(n)}+\frac{k-1}{\log _{2}(n)}\Bigg). 
\end{equation} 
 
\item (iii) The algorithm $LZ77$ is 8-asymptotically optimal with 
respect to $\hat{H}_{1}$ and for any string $\omega ^{n}$ such that 
$\hat{H}_{1}(\omega ^{n})\neq 0$ it holds 
\begin{equation} 
I_{LZ77}(\omega ^{n})\leq \ 8\ n\ \hat{H}_{1}(\omega ^{n})+\ O\Big(t\log 
_{2}\log _{2}\Big(\frac{n}{t}\Big)\Big)\ .  \label{lz8opt} 
\end{equation} 
\end{description} 
\end{theorem} 
\subsection{$LZ77$ on the Manneville map}\label{man77} 
Now we are ready to prove the following theorem, which links the 
Information Content obtained via the algorithm $LZ77$ to the $AIC$ on 
the symbolic orbits of the Manneville map. We will study the dynamical 
system $([0,1],\mathcal{T}_z)$ where $\mathcal{T}_z (x)=x+x^z (mod \ 
1)$ and $z>1$. The reference measure is the Lebesgue measure on the 
unit interval. 
 
The Manneville map was introduced by P. Manneville in 
\cite{Manneville} as an example of a discrete dissipative dynamical 
system with \textit{intermittency}: there is an alternation between 
long regular phases, called \textit{laminar}, and short irregular 
phases, called \textit{turbulent}. This behavior has been observed 
in fluid dynamics experiments and in chemical reactions. 
 
In order to state and prove our results, we recall some useful lemmas 
coming from pro\-ba\-bi\-li\-ty theory. 
 
\begin{lemma} 
\textbf{(Jensen's Inequality)} Let $I$ be a closed interval in 
$\mathbb{R}$ and\linebreak $u:I\longrightarrow \mathbb{R}$ be convex 
and continuous at the endpoints of $I$. If $X$ is a random variable 
which takes its values in $I$, then $\mathbb{E}[u\circ X]\geq 
u(\mathbb{E}[X])\ .$ 
\end{lemma} 
 
\begin{lemma} 
\label{lemmax} If $X$ and $Y$ are two real random variables s.t. $X\geq 0$ 
and $Y\geq 0$, then  
\begin{equation*} 
\mathbb{E}[\max \{X,Y\}]\geq \max \{\mathbb{E}[X],\mathbb{E}[Y]\}\ . 
\end{equation*} 
\end{lemma} 
\begin{theorem}\label{41} 
\label{tman77}  
Consider the dynamical system $([0,1],\mathcal{T}_z)$ driven by the 
Manneville map $\mathcal{T}_z (x)=x+x^z (mod \ 1)$, with $z>1$. Let 
$\tilde{x}\in (0,1)$ be such that $\mathcal{T}_z(\tilde{x})=1$. 
Consider the partition $\alpha=\{[0,\tilde{x}],(\tilde{x},1]\}$ 
of the unit interval $[0,1]$.  If $\omega $ is a symbolic orbit drawn 
from the Manneville map, with respect to the partition $\alpha$, 
then 
\begin{equation} 
\begin{array}{cr} 
\mathbb{E}[I_{LZ77}(\omega ^{n})]\sim n & \mbox{if }z<2 \\[10pt] 
O(n^{p })\leq \mathbb{E}[I_{LZ77}(\omega ^{n})]\leq O(n^{p }\log 
_{2}(n)) & \mbox{if }z>2 
\end{array} 
\end{equation} 
where $p =\frac{1}{z-1}$ and the measure is the usual Lebesgue measure 
on the interval. 
\end{theorem} 
 
\begin{proof} 
We will study the two cases separately. 
 
\begin{description} 
\item  {If $z<2$: } a result of \cite{GW} shows that for the expectation 
value of the $AIC$ of a symbolic orbit of the Manneville map with $z<2$, 
with respect to the Lebesgue measure on the interval, it holds that  
\begin{equation*} 
\mathbb{E}[AIC(\omega ^{n})]\sim n\ . 
\end{equation*} 
Since it is $AIC(\omega ^{n})\leq I_{LZ77}(\omega ^{n})\leq O(n)$, then in 
this case we have that $\mathbb{E}[I_{LZ77}(\omega ^{n})]\sim n$. 
 
\item  {If $z>2$: } From Theorem \ref{teomanz} $(iii)$, we know that the 
algorithm $LZ77$ is asymptotically optimal with respect to $\hat{H}_{1}$, 
that is  
\begin{equation} 
I_{LZ77}(\omega ^{n})\leq \ 8\ n\ \hat{H}_{1}(\omega ^{n})\ +\ O\Big(t\log 
_{2}\log _{2}\Big(\frac{n}{t}\Big)\Big)\ , 
\end{equation} 
where $t$ is the number of words in the $LZ77$ parsing of $\omega ^{n}$. 
Thus, for any sequence $\omega ^{n}$, it holds  
\begin{equation} 
\mathbb{E}[I_{LZ77}(\omega ^{n})]\leq \ 8\ n\ \mathbb{E}[\hat{H}_{1}(\omega 
^{n})]\ +\ \mathbb{E}\Big[t\log _{2}\log _{2}\Big(\frac{n}{t}\Big)\Big]\ . 
\label{dadim} 
\end{equation} 
First, we will prove that $n\ \mathbb{E}[\hat{H}_{1}(\omega ^{n})]$ is 
bounded by $O(n^{p }\log _{2}(n))$ with $p =\frac{1}{z-1}$. Then, 
we will give an estimate for $\mathbb{E}[t\log _{2}\log _{2}(\frac{n}{t})]$, 
so completing the proof. 
 
By definition, $\hat{H}_{1}(\omega ^{n})=-(\frac{N_{n}}{n}\log 
_{2}(\frac{ N_{n}}{n} )+(1-\frac{N_{n}}{n})\log 
_{2}(1-\frac{N_{n}}{n}))$ where $N_{n}$ is the number of occurrences 
of the event $\mathcal{E}=\{passage\ through\ (\tilde{x} ,1]\}$. 
 
In \cite{feller} it has been proved that if $z>2$ then $\mathbb{E} 
[N_{n}]\sim n^{p }$ . 
 
Therefore, for the first order empirical entropy of a symbolic orbit drawn 
by the Manneville map with respect to the Lebesgue measure, we can apply 
Jensen's inequality and obtain: 
 
\begin{equation*} 
\begin{array}{ll} 
\mathbb{E}[\hat{H}_{1}(\omega ^{n})] & \leq  - 
\Bigg(\mathbb{E}\Bigg[\frac{ N_{n} }{n}\log 
_{2}\Big(\frac{N_{n}}{n}\Big)\Bigg]+\mathbb{E}\Bigg[\Big(1- \frac{ 
N_{n}}{n}\Big)\log _{2}\Big(1-\frac{N_{n}}{n}\Big)\Bigg]\Bigg)\leq \\ 
[15pt] & \leq  - \Bigg(\mathbb{E}\Big[\frac{N_{n}}{n}\Big]\ \log 
_{2}\Big(\mathbb{ \ E}\Big[\frac{N_{n}}{n}\Big]\Big)+ \\[15pt] & 
\qquad \qquad \qquad \qquad \quad 
+\Big(1-\mathbb{E}\Big[\frac{N_{n}}{n} \Big]\Big)\ \log 
_{2}\Big(1-\mathbb{E}\Big[\frac{N_{n}}{n}\Big]\Big)\Bigg) \sim 
\\[15pt] & \sim  - \Bigg(n^{p -1}\log _{2}\Big(n^{p -1}\Big)+\Big( 
1-n^{p -1}\Big)\log _{2}\Big(1-n^{p -1}\Big)\Bigg)\ . 
\end{array} 
\end{equation*} 
Consequently, we can easily verify that  
\begin{equation*} 
\begin{array}{ll} 
8\ n\ \mathbb{E}[\hat{H}_{1}(\omega ^{n})] & \leq \ -\ 8\ n^{p }\log 
_{2}\Big( n^{p -1}\Big)-8\ \Big(n-n^{p }\Big)\log _{2}(1-n^{p 
-1} )= \\[15pt] & =\ 8\ n^{p }\log _{2}\Big(n^{1-p }-1\Big)-\ 8\ n\ 
\log _{2}\Big( 1-n^{p -1} \Big)\ . 
\end{array} 
\end{equation*} 
For the fact that $p <1$, it holds  
\begin{equation} 
\begin{array}{l} 
\ 8\ n^{p }\log _{2}\Big(n^{1-p }-1\Big)-\ 8\ n\ \log _{2}\Big(  
1-n^{p -1}\Big)\sim \\[15pt]  
\sim \ \ n^{p }\Big((1-p )\log _{2}(n)-1\Big)\ = \\[15pt]  
=\ O(n^{p }\log _{2}(n))\ . 
\end{array} 
\label{stimah0} 
\end{equation} 
Now we will prove that, for $\omega ^{n}$ a symbolic orbit of the Manneville 
map with $z>2$, if $t$ is the number of words in the $LZ77$ parsing of $ 
\omega ^{n}$, it holds  
\begin{equation} 
\mathbb{E}\Big[t\log _{2}\log _{2}\Big(\frac{n}{t}\Big)\Big]\leq O(n^{p 
}\log _{2}(n))\ .  \label{stimat} 
\end{equation} 
We apply Lemma \ref{lemmat}, with alphabet $\mathcal{A}=\{0,1\}$ where 
$1$ is the symbol associated to the event $\mathcal{E}=\{passage\ 
through\ [\tilde{x} ,1]\}$, which appears in $\omega ^{n}$ with mean 
probability $\frac{N_{n}}{n} =n^{p -1}$ and $0$ is the symbol 
associated to the event $not\ \mathcal{ \ E}$. 
 
Thus, $m=n-\max \{n_{0},n_{1}\}$ and $t\leq 2(m+1)=2(n-\max 
\{n_{0},n_{1}\}+1)\ .$ 
 
Thanks to Lemma \ref{lemmax}, we obtain the following estimates:  
\begin{equation} 
\begin{array}{ll} 
\mathbb{E}[t] & \leq \ 2\ \mathbb{E}[r]+1= \\  
& =\ 2\ n+1-2\ \mathbb{E}[\max \{n_{0},n_{1}\}]\leq \\  
& \leq \ 2\ n+1-2\ \max \{\mathbb{E}[n_{0}],\mathbb{E}[n_{1}]\}= \\  
& =\ 2\ n+1-2\ \max \{n-n^{p },n^{p }\}= \\  
& =\ 2\ n^{p }+1\ . 
\end{array} 
\end{equation} 
Eventually, the inequality (\ref{stimat}) can be easily verified. 
 
From the estimates (\ref{stimah0}) and (\ref{stimat}) together with (\ref 
{dadim}), it follows that  
\begin{equation*} 
\mathbb{E}[I_{LZ77}(\omega ^{n})]\leq O(n^{p }\log _{2}(n))\ . 
\end{equation*} 
Finally, since the $AIC$ is the ideal information content of a string, it is  
$I_{LZ77}(\omega ^{n})\geq AIC(\omega ^{n})$ and the same inequality relates 
the expectation values. 
 
In \cite{claudio} and \cite{gal3}, it has been proved that $\mathbb{E} 
[AIC(\omega ^{n})]\geq O(n^{p })$. This completes the proof. 
\end{description} 
\end{proof} 
 
\subsection{The algorithm CASToRe} 
 
We have created and implemented a particular compression algorithm we 
called CASToRe which is a modification of the well known LZ78 
algorithm (\cite{lz78}). Its theoretical advantages with respect to 
LZ78 are shown in \cite{ChaSolFra}, \cite{mb}: it is a sensitive 
measure of the Information content of low entropy sequences. That's 
why is called \textbf{CASToRe}: \textbf{C}ompression 
\textbf{A}lgorithm, \textbf{ S}ensitive \textbf{To} 
\textbf{Re}gularity. 

As it has been proved in Theorem 4.1 in \cite{mb}, the Information $I_Z$
of a constant sequence $s^n$, originally with length $n$, is $4+2\log
(n+1)[\log (\log (n+1))-1]$, if the algorithm $Z$ is CASToRe. The
theory predicts that the best possible information for a constant
sequence of length $n$ is $AIC(s^n) =\log (n) + $const. In
\cite{ChaSolFra}, it is shown that the algorithm $LZ78$ encodes a
constant $n$ digits long sequence to a string with length about
$const\ +\ n^{\frac 1 2}$ bits; so, we cannot expect that $LZ78$ is
able to distinguish a sequence whose information grows like
$n^{\alpha}$ ($\alpha < \frac 1 2$) from a
constant or periodic one. This motivates the choice of using CASToRe.

Now we briefly describe the internal running of CASToRe.

As the algorithm LZ77, the algorithm CASToRe is based on an adaptive 
dictionary (\cite{bell}). One of the basic differences in the coding 
procedure is that the algorithm LZ77 splits the input strings in 
overlapping words, while the algorithm CASToRe (as already the 
algorithm LZ78) parses the input string in non-overlapping words. 
 
At the beginning of encoding procedure, the dictionary is contains
only the alphabet. In order to explain the principle of encoding,
let's consider a step $h$ within the encoding process, when the
dictionary already contains $h$ words $\{e_1,\dots,e_h\}$.
 
The new word is defined as a pair ({\it prefix pointer},{\it suffix 
pointer}). The two pointers are referred to two (not necessarily 
different) words $\rho_p$ and $\rho_s$ chosen among the ones contained 
in the current dictionary as follows. First, the algorithm reads the 
input stream starting from the current position of the front end, 
looking for the longest word $\rho_p$ matching the stream. Then, we 
look for the longest word $\rho_s$ such that the joint word $\rho_p 
\rho_s$ matches the stream. The new word $e_{h+1}$ which will be added 
to the dictionary is then $e_{h+1}=\rho_p \rho_s$. 
 
The output file contains an ordered sequence of the binary encoding of 
the pairs $(i_p,i_s)$ such that $i_p$ and $i_s$ are the dictionary 
index numbers corresponding to the prefix word $\rho _p$ and to the 
suffix word $\rho_s$, respectively.  The pair $(i_p,i_s)$ is referred 
to the new encoded word $e_{h+1}$ and has its own index number 
$i_{h+1}$. 
 
 The following example shows how the algorithm CASToRe encodes the 
input stream 
\begin{equation*} 
\omega =(abcababccabb\dots) . 
\end{equation*} 
 
Let the source alphabet be $\mathcal{A}=\{a,b,c\}$. 
 
The output is the binary encoding of the following pairs contained in 
the second column. The first column is the dictionary index number of 
the encoded word in the dictionary which is showed in the same line, 
second column. For an easier reading, we add a third column which 
shows each encoded word in the original stream $\omega$, but it is not 
contained in the output file: 
$$ 
\begin{array}{lll} 
&\mbox{First, the dictionary is being loaded}&\\
1 & (0,\ ^{\prime}a\ ^{\prime}\ ) & [a] \\  
2 & (0,\ ^{\prime}b\ ^{\prime}\ ) & [b] \\  
3 & (0,\ ^{\prime}c\ ^{\prime}\ ) & [c] \\  
&\mbox{Then, the encoding procedure starts}&\\
4 & (1,2) & [ab] \\  
5 & (3,4) & [cab] \\  
6 & (4,3) & [abc]\\
7 & (5,3) & [cabc]
\end{array} 
$$ 
and so on. 
 
We remark that this coding procedure, which pairs words already in the 
dictionary to create a new word, is similar to the procedure that can 
be found in the recent work \cite{grass}, which seems to be able to 
give a very precise entropy estimation, detecting very long range 
correlations in the English language. 
 
\section{Numerical experiments} 
 
In this section we show some numerical experiments supported by the 
theory of the previous sections. We consider some examples of fully 
chaotic and weakly chaotic dynamical systems and we measure the 
information content of the generated symbolic orbits with respect to 
some partition.  
 
We measure the information content with the two different data 
compression algorithms LZ77 with infinite window and CASToRe, whose 
internal running has been presented above. These two algorithms seems 
to be suitable for the compression of 0-entropy strings (see 
\cite{ChaSolFra} and Proposition \ref{teomanz}) and they are fast 
enough to allow the compression of long strings (we could manage to 
compress trajectories of $O(10^7)$ symbols). From the computational 
point of view, whereas LZ77 requires a big amount of RAM (Random 
Access Memory), since it needs to retain the entire string already 
encoded and the entire dictionary built, the algorithm CASToRe only 
remembers the dictionary which is implemented in a tree 
structure. Hence, the computation time is a distinguishing feature 
between the two algorithms: CASToRe can compress a $O(10^7)$-symbols 
string in few seconds. 
 
As we will see the results agree with theoretical predictions when 
they are available or with other numerical results which can be found 
in the literature. 
 
It is worth to remark that, even if the two compression schemes are 
basically different, the two algorithms give a behavior of the 
information content of the same order in all the numerical experiments 
we performed. 
 
\subsection{The Manneville map}\label{numMan} 
 
We measure the information content of symbolic orbits drawn from the 
Manneville map. 
 
Let us consider again the Manneville map $\mathcal{T}_z (x)=x+x^z (mod 
\ 1)$ as it has been presented in Section \ref{man77}. 
 
Let us consider the partitions $\alpha _1, \alpha_2 $ obtained by 
dividing $[0,1]$ in 2 or 4 subintervals. The partition $\alpha_1$ is 
the same described in Section \ref{man77} and $\alpha_2$ is a 
refinement of $\alpha_1$: $\alpha_2$ is obtained splitting in two 
equal parts each interval of $\alpha_1$. We denote by $K(x,\alpha 
_i,n), i \in\{1,2\}$ the Algorithmic Information Content of a $n$-long 
symbolic orbit of the Manneville map with initial condition $x$, with 
respect to the partition $\alpha _i$. 
 
\begin{figure}[h] 
\begin{tabular}{lr} 
{\raggedright{ 
\psfig{figure=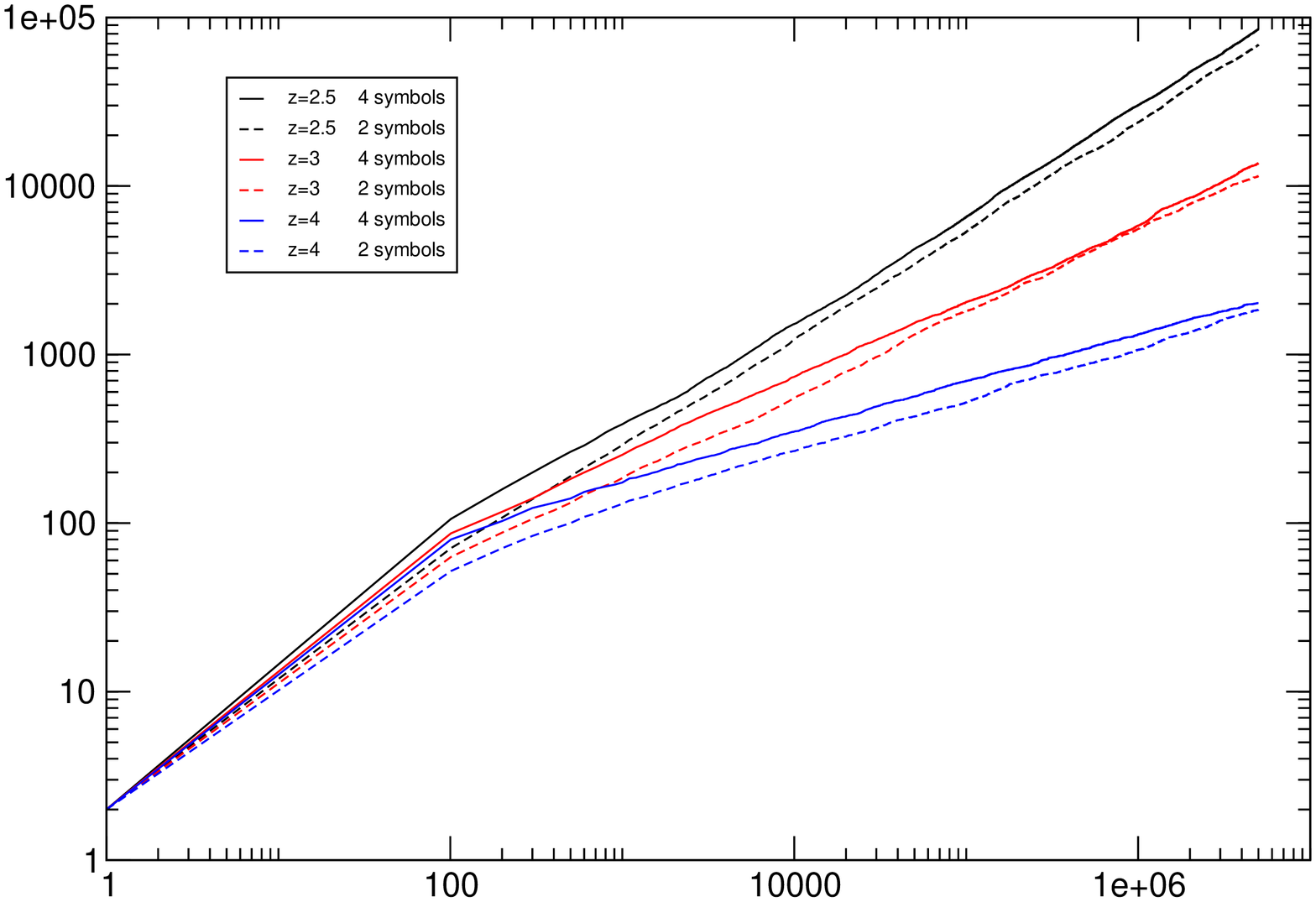,width=6.4cm,angle=270}}} & { 
\raggedleft{ 
\psfig{figure=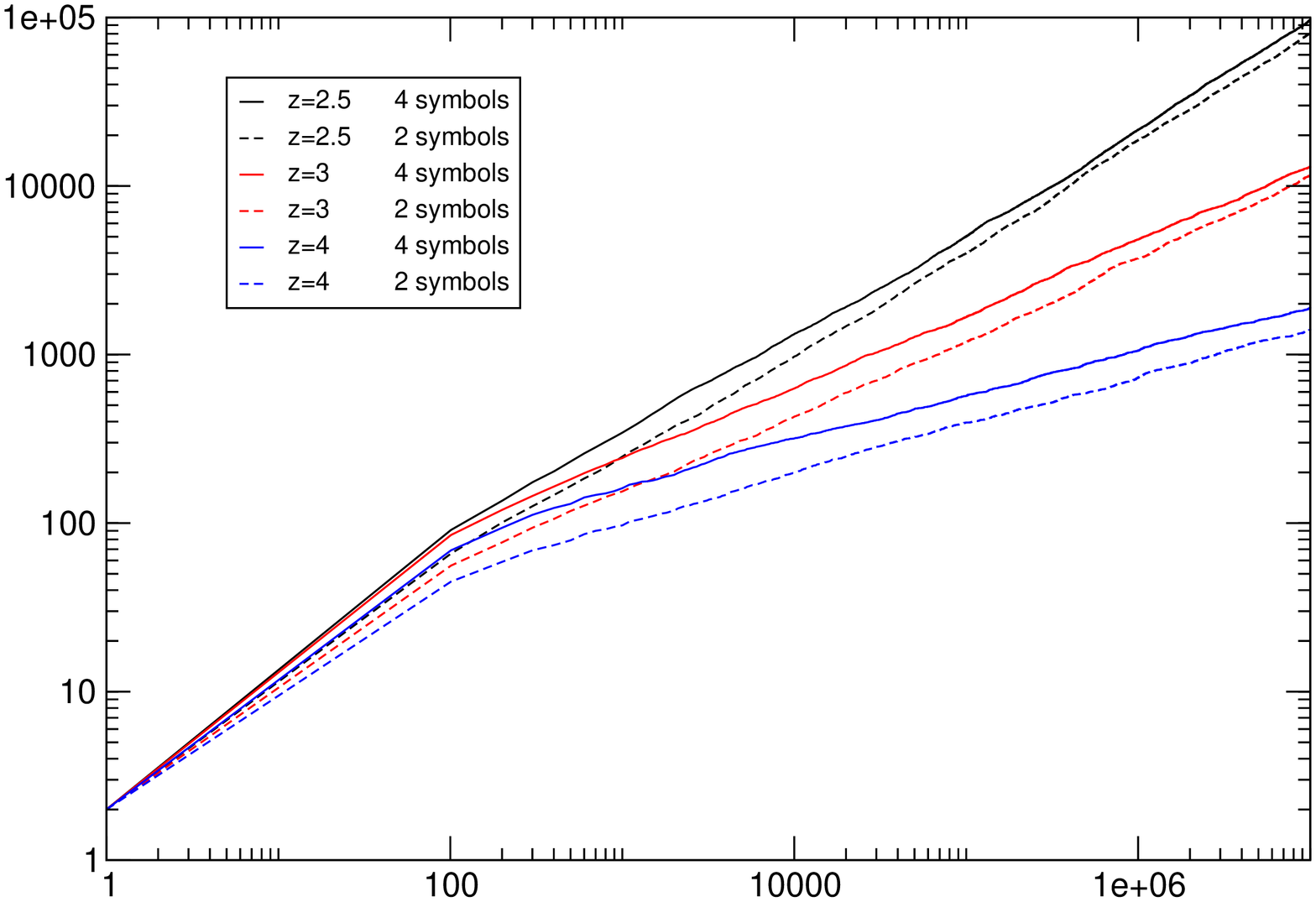,width=6.4cm,angle=270}}} 
\end{tabular} 
\caption{\it The plot of the above experiment: each graph represents 
the behavior of the information plotted versus the number of steps (in 
log-log scale). Note that in log-log scale power laws become straight 
lines. On the left we plot the information as it is measured by 
CASToRe, on the right we plot the information as it is measured by 
LZ77. } 
\label{figmann} 
\end{figure}

By the results exposed in (\cite{GW},\cite{claudio}, \cite{gal3}) we 
have that the mean value of $K(x,\alpha _i,n)$, with respect to the 
Lebesgue measure, on the initial conditions of the orbit is expected 
to be $\mathbb{E}[K(x,\alpha _i,n)]\sim n^{p}$, with $p=\frac{1}{z-1}$ for 
$z>2$, and $\mathbb{E}[K(x,\alpha _i,n)]\sim n$ for $z<2$.  Moreover (by 
the results of section \ref{man77}) the same result holds for the information 
as it is measured by the algorithm LZ77. We verified numerically this 
statement, and the result is shown in figure \ref{figmann}. 
 
If we use the information content as it is measured by CASToRe 
the numerical result is also close to the previous one. This confirms 
the theoretical results and proves that the methods relative to the 
Computable Information Content are experimentally reliable. 
 
We considered a set of one hundred initial points, then we generated 
the relative $10^{7}$-long orbits and we applied the compression 
algorithms to the associated symbolic strings $s$ (with respect both 
to the partition $\alpha _1$ and $\alpha _2$). 
 
In Table \ref{tabellamann} we show the results. The first column 
indicates the partition to which the results are referred. The second 
column is the value of the parameter $z$ which drives the dynamics of 
the system. The last column gives the results of the theory for the 
exponent $p$ of the asymptotic behavior of $K(x,\alpha _i,n)\sim 
n^{p}$. The third and the fourth columns show the experimental 
results. The shown number is the average $\bar p$ of the exponents of 
one hundred different orbits. The initial conditions of the orbits are 
chosen randomly with respect to the Lebesgue measure. 
 
\begin{center} 
\begin{table}[th] 
\begin{tabular}{|c|c|c|c|c|} 
\hline 
Symbols & z & LZ77 & CASToRe & Theoretical value \\ \hline 
4 & 2.5 & 0.64 & 0.64 & 0.66 \\ \hline 
2 & 2.5 & 0.64 & 0.64 & 0.66 \\ \hline 
4 & 3 & 0.49 & 0.43 & 0.5 \\ \hline 
2 & 3 & 0.47 & 0.48 & 0.5 \\ \hline 
4 & 4 & 0.27 & 0.25 & 0.33 \\ \hline 
2 & 4 & 0.32 & 0.28 & 0.33 \\ \hline 
\end{tabular} 
\caption{\it Theoretical and experimental results for the Information 
content of the Manneville map} 
\label{tabellamann} 
\end{table} 
\end{center} 
 
In Figure \ref{figmann} are plotted several examples of the behavior 
of the Information Content $I_Z$ when $Z=$LZ77 (on the right) or 
$Z=$CASToRe (on the left) and for different values of the parameter 
$z$. The scale is bilogarithmic, so that the power laws become 
straight lines and the exponent $p$ of the expected power law is the 
slope of the correspondent straight line. 
 
\subsection{The logistic map\label{logistic}} 
 
In this section we study the logistic map at the chaos threshold from 
an ex\-pe\-ri\-men\-tal point of view.  We recall that the logistic 
map is defined by 
\begin{equation} 
f(x)=\lambda x(1-x)\ ,\quad x\in [0,1]\ ,\quad1\leq\lambda\leq4. 
\label{logmap} 
\end{equation} 
 
The logistic map has been used to simulate the behavior of biological 
species not in competition with other species. Later the logistic map 
has also been presented as the first example of a relatively simple 
map with an extremely rich dynamics (\cite{collet},\cite{feigen}). If 
we let the parameter $\lambda$ vary from $1$ to $4$, we find a 
sequence of bifurcations of different kinds. For values of $\lambda 
<\lambda_{\infty}=3.56994567187\dots$, the dynamics is periodic and 
there is a sequence of period doubling bifurcations which leads to the 
chaos threshold for $\lambda=\lambda_{\infty}$.

Numerical experiments and heuristics considerations from the physical 
literature indicate that at the chaos threshold there is a power-law 
``sensitivity'' to initial conditions (here the sensitivity to initial 
conditions was measured with a generalized Lyapunov exponent). These 
facts justified the application of generalized entropies to the map 
(\cite{tsallis}). 
 
Moreover by the more recent results of \cite{mb} we know that if we 
consider the Lebesgue measure, then for almost any initial condition 
the Algorithmic Information Content of an orbit will increase as the 
logarithm of the number of steps. 
 
In the following, we will show how we have experimentally confirmed 
this result measuring the information with LZ77 and CASToRe. 
 
\begin{figure}[h] 
\begin{tabular}{lr} 
{\raggedright{\psfig{figure=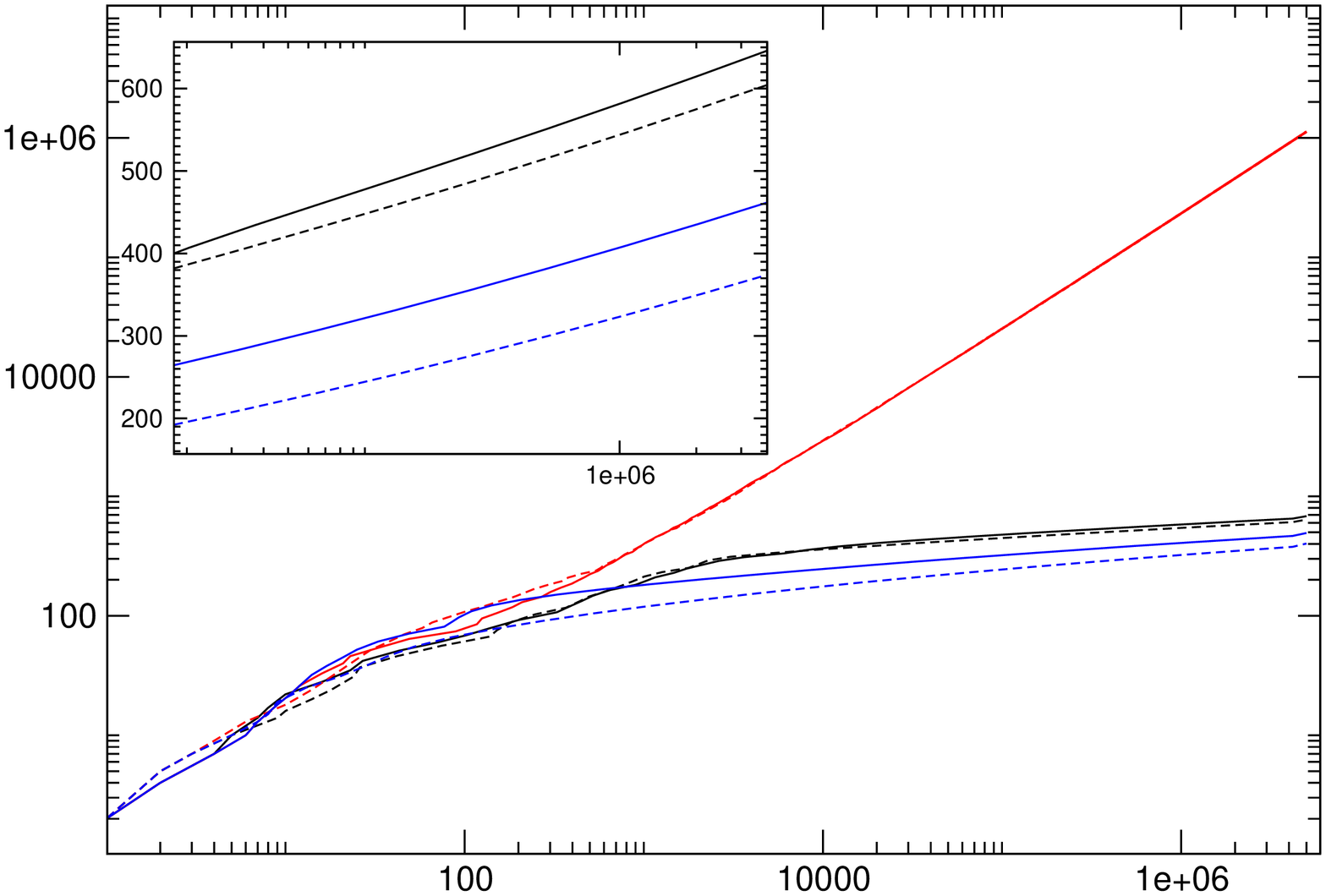,width=6.4cm,angle=270} 
}} & {\raggedleft{ 
\psfig{figure=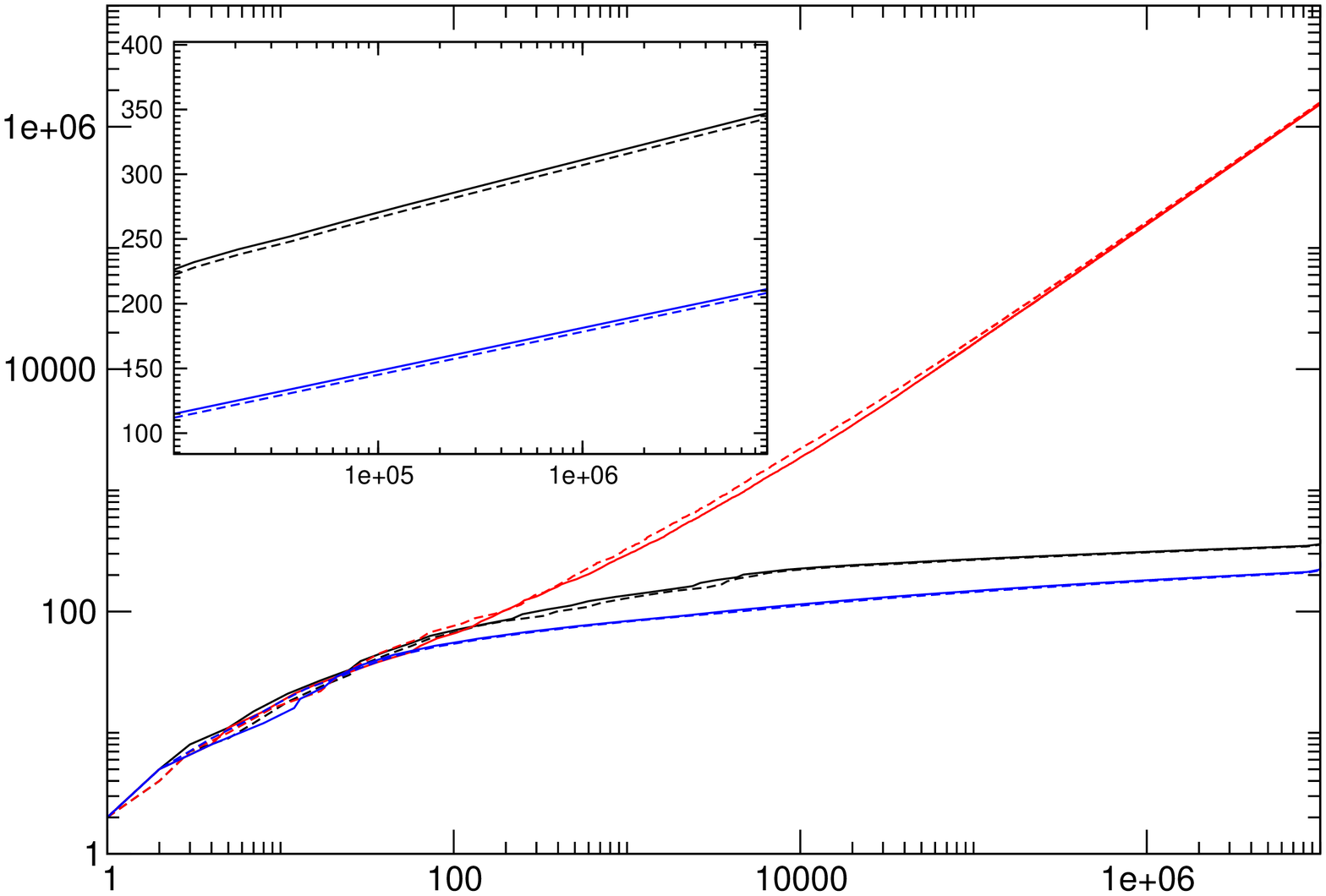,width=6.4cm,angle=270}}} 
\end{tabular} 
\caption{\it On the left we have the information v.s. number of steps for a 
typical point of the interval as it is measured by CASToRe, on the right the 
same with LZ77. The plot is in log-log scale, while in the inset  
the plot is log-linear to show how the long time behavior of the information 
follows a logarithmic increase. } 
\label{figlogistica} 
\end{figure} 
 
In figure \ref{figlogistica} the main plot is in bilog scale, while 
the inset is in log-linear scale and the same graphs as in the main 
plot are pictured. On the left, the experiments performed via CASToRe, 
on the right via LZ77. The analysis of results is the same for both 
pictures. 
 
The solid line in the main plot represents the information behavior 
at the chaos threshold.  This graph already indicates that at the 
chaos threshold $\lambda_{\infty}$ the information increases below any 
power law (any power law becomes a straight line when plotted in bilog 
scale and our graph is evidently concave), as predicted by the 
theory. A more accurate quantitative analysis was done in \cite{mb}. 
The upper lines are referred to values of $\lambda >\lambda_{\infty}$ 
for which the map is chaotic (the information increases linearly with 
time). The lower lines represents the information behavior when 
$\lambda$ tends to $\lambda_{\infty}$ from below (along the period 
doubling cascade), hence when we are in the periodic regime (where the 
Algorithmic Information Content is expected to behave 
logarithmically). 
 
\subsection{Tirnakli, Tsallis, Lyra (TTL)-circular like maps} 
 
These maps have been introduced in \cite{TTL}, where they are studied 
numerically, as modifications of the classical standard map. These 
maps are one-dimensional maps and varying the parameters they show a 
transition to chaos. They are defined by 
\begin{equation*} 
T_{z}(x)=\Omega _z + (x-\frac{1}{2\pi} \ \sin (2\pi x ))^{\frac z 3} 
(mod \ 1) 
\end{equation*} 
and we study the maps with parameters values $z=3,z=4,z=5$ with 
$\Omega_3 = 0.606661063469$, $\Omega_4 = 0.648669091983$, $\Omega _5= 
0.6788311756505$, for which values the maps are at the onset of chaos. 
 
We recall that for $z=3$ we obtain the classical the standard map 
\begin{equation*} 
T(x)=\Omega + (x-\frac{K}{2\pi} \ \sin(2\pi x )) (mod \ 1) 
\end{equation*} 
with $K=1$ (at the edge of a quasiperiodic transition to chaos). 
 
\begin{figure}[h] 
\centerline{\psfig{figure=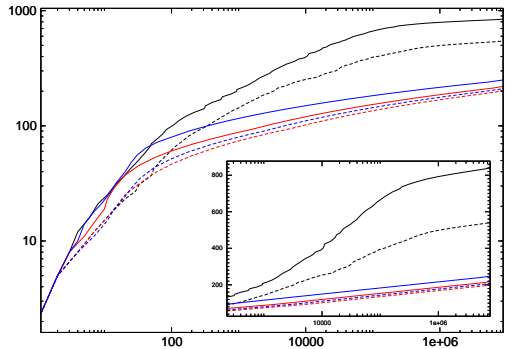,width=8cm,angle=270}} 
\caption{\it The information vs. number of steps for the TTL-circular
like maps for a typical point studied using the algorithm CASToRe. The
solid line is referred to the map for z=1 with 2 or 4 symbols, fr the
dashed and dotted curves, we have z=4 and z=5. Inset: same graph in
log-linear scale.}
\label{figtsallis} 
\end{figure}

For these maps results in \cite{TTL} show a numerical evidence of
power law initial data sensitivity, as it was shown in \cite{tsallis}
for the logistic map at the edge of chaos. Also, by the results of
\cite{gal3}\footnote{In the cited paper, quantitative results are
proved between initial condition sensitivity and complexity.}, this
would correspond to a logarithmic increase of the algorithmic information. We
measured the information coming from these maps, obtaining a behavior
that is also similar to the logistic map at the edge of chaos and fits
with the cited numerical results (Figure \ref{figtsallis}).
 
\subsection{Casati-Prosen map} 
 
This area-preserving map has been proposed in \cite{casati} as a model 
of quantum chaos. The map is defined on $T^2= [-1,1)\times[-1,1)$ by 
$T \left( 
\begin{array}{c} 
x_{n} \\  
y_{n} 
\end{array} 
\right)=\left( 
\begin{array}{c} 
x_{n+1} \\  
y_{n+1} 
\end{array} 
\right)$ where  
\begin{equation*} 
\begin{array}{cl} 
x_{n+1} \ = & x_n+y_{n+1} \\[0.3cm] 
y_{n+1} \ = & y_{n}+\alpha \mbox{ sgn}(x_{n}) +\beta  
\end{array} 
\end{equation*} 
and $\alpha=\frac{ (\frac 12 (\sqrt{5}-1)-\frac 1e)}2$, $\beta 
=\frac{(\frac 12 (\sqrt{5}-1)+\frac 1e)}2$. Results in \cite{casati} 
provide numerical evidences that the map is ergodic and mixing, with 
linear speed of separation of nearby starting orbits. 
 
We studied the complexity of some trajectory of the system, obtaining
that the computable information seems to increase as a power law $n^p$ with
exponent $p$ approximately equal to $0.75\dots$ ($p= 0.742$ when
estimated by LZ77 and $p= 0.755$ when estimated by CASToRe). This
result is quite unexpected from the connection between sensitivity to
initial conditions and the asymptotic behavior of the information
content (\cite{gal3}). From the cited results, to a linear initial
condition sensitivity would correspond a logarithmic increase of the
AIC. However, the rigorous proof of all the properties of the map
remains an open problem.
 
\begin{figure}[h] 
\begin{tabular}{lr} 
{\raggedright{\psfig{figure=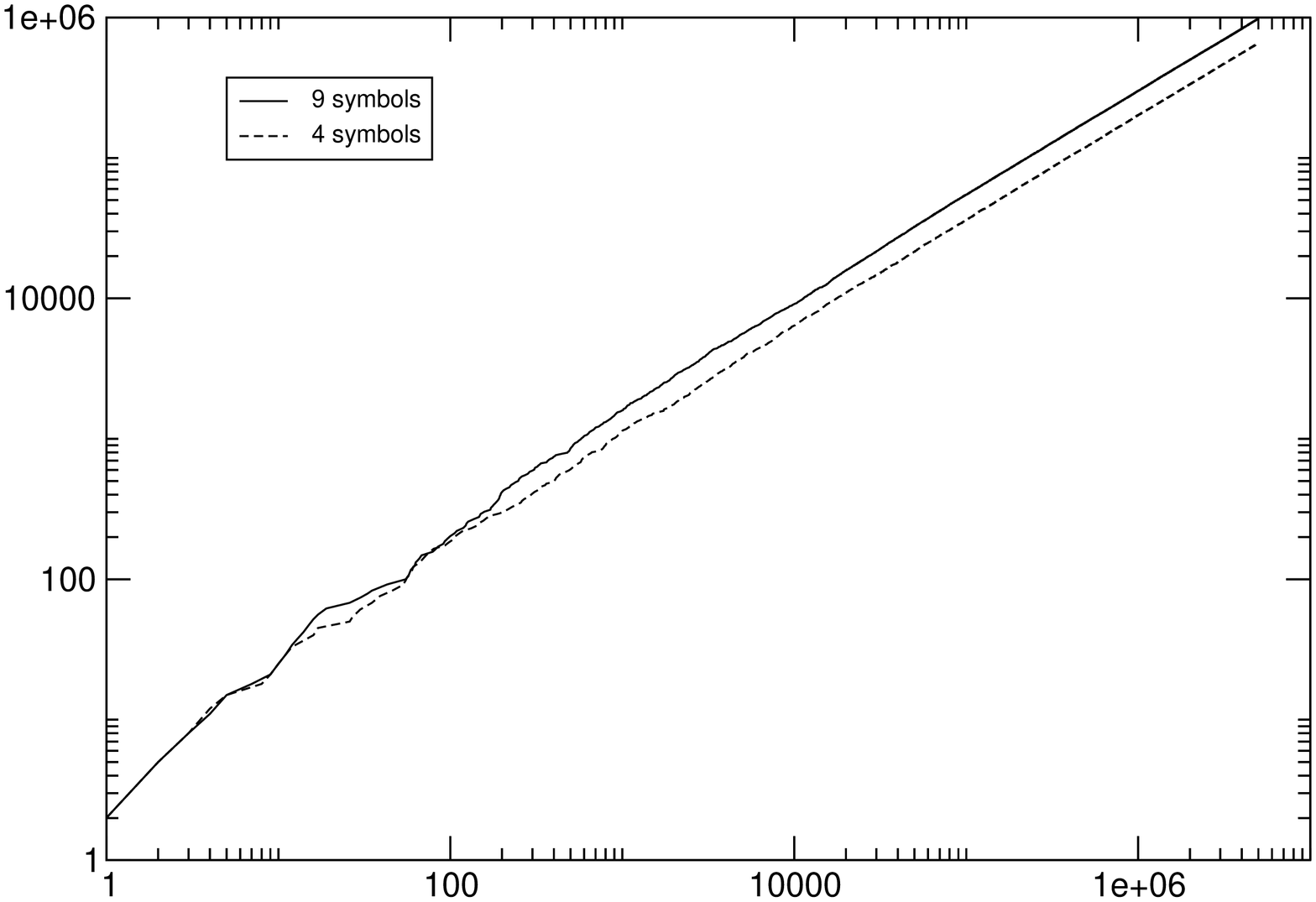,width=6.4cm,angle=270}}} 
& {\raggedleft{ 
\psfig{figure=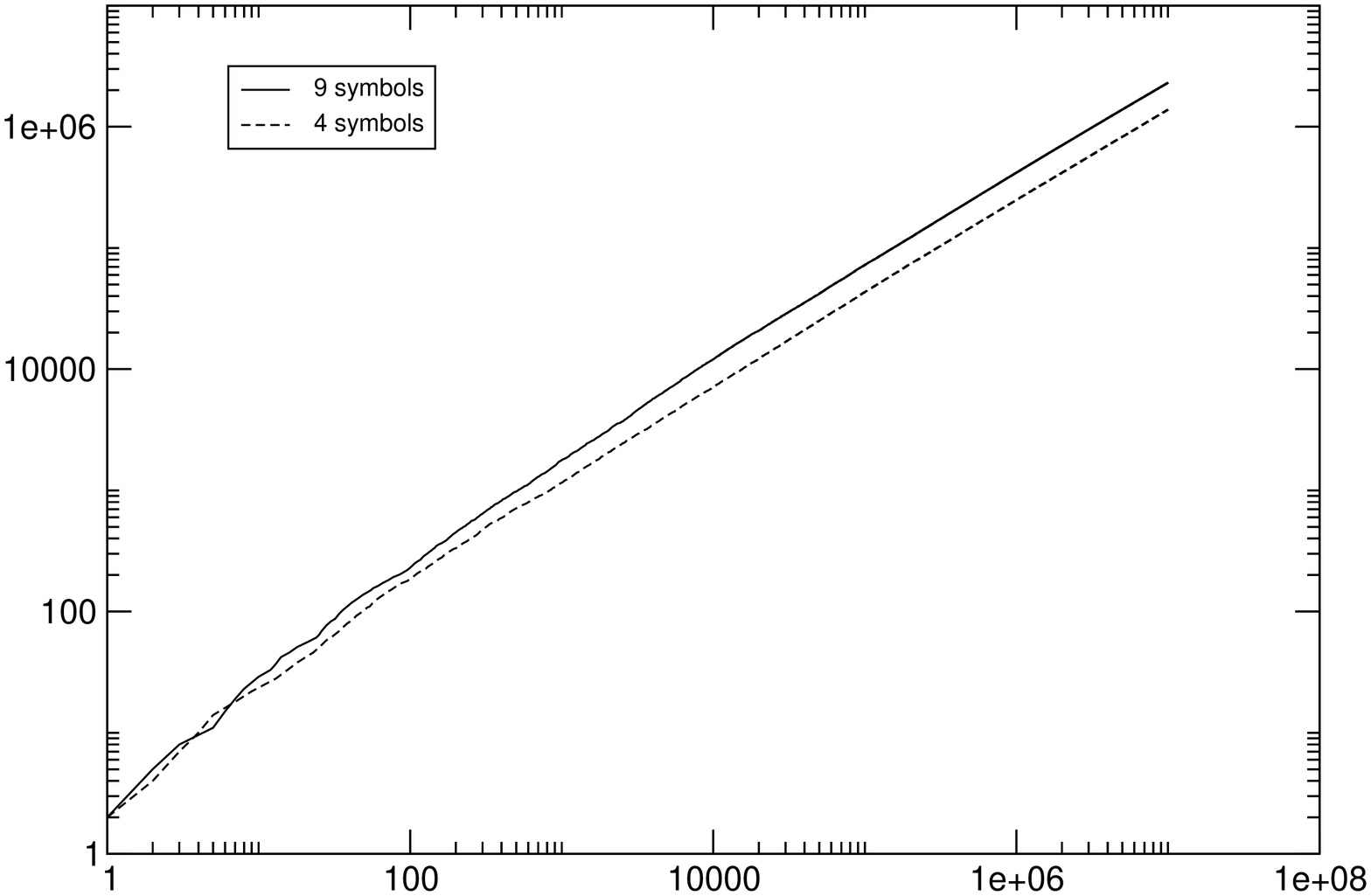,width=6.4cm,angle=270}}} 
\end{tabular} 
\caption{\it The information vs. number of steps (for CASToRe, on the 
left, and LZ77, on the right) for a typical point in the Casati-Prosen 
map. The plot is in log-log scale.} 
\label{figcasati} 
\end{figure}

\subsection{The Arnold cat map} 
 
\begin{figure}[ht] 
\centerline{\psfig{figure=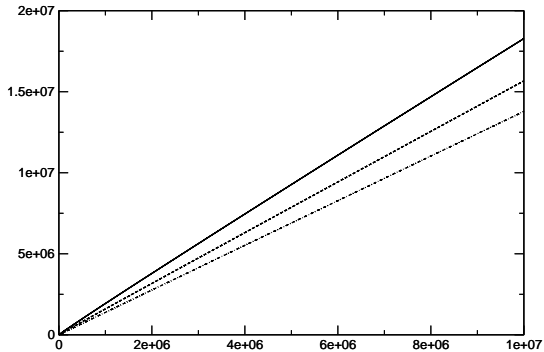,width=8cm,angle=270}} 
\caption{\it The information content of the Arnold cat map for three 
different partitions. The dashed line is the information content of 
the generating partition. For the partitions used see the description 
in the text.} 
\label{figcat} 
\end{figure}

The Arnold cat map is an example of a two-dimensional hyperbolic toral 
automorphism, that is the projection on the two-torus 
$\mathbb{R}^2/\mathbb{Z}^2$ of a linear map of $\mathbb{R}^2$, 
represented by a matrix $M$ with integer elements and determinant one 
and real eigenvalues $\lambda$ and $1/\lambda$, different from 1. The 
Arnold cat map is specified by the matrix 
$$M= \left( 
\begin{array}{cc} 
1 & 1 \\ 
1 & 2  
\end{array} 
\right)$$ with $\lambda=(3+\sqrt{2})/2$. From a theorem of 
\cite{arn-avez} it follows that the Kolmogorov-Sinai entropy with 
respect to the Lebesgue measure of a two-dimensional hyperbolic toral 
automorphism is given by the logarithm of the modulus of the 
eigenvalue bigger than 1. Hence, in this case $$h = \log_2 
\frac{3+\sqrt{2}}{2} \sim 1.388\dots$$ Our computations give the same 
result. We only show the results for the compression algorithm 
CASToRe, since in this case no evident differences can be 
appreciated. Studying the behavior of the information content with 
respect to the length of the compressed string, we expect to find a 
straight line with angular coefficient equal to the entropy of the 
dynamical system, when working with a generating partition. In figure 
\ref{figcat} it is represented the information function for three 
different choices of the partition. The dotted line is obtained with a 
partition of the square $[0,1)\times [0,1)$ in two horizontal 
strips. The solid line is obtained with a partition in four equal 
squares along the axis, and the dashed line is obtained with a 
partition in four squares along the eigen-directions associated to the 
matrix $M$. The angular coefficients of the lines are 0.98 for the 
dotted line, 1.56 for the solid line and 1.37 for the dashed line, 
showing that the first two partitions considered are not able to 
simulate the whole complexity of the system, whereas the last one can 
be considered to be a generating partition. 
 
\subsection{The Froeschl\'e map} 
\begin{figure}[ht] 
\begin{tabular}{lr} 
{\raggedright{\psfig{figure=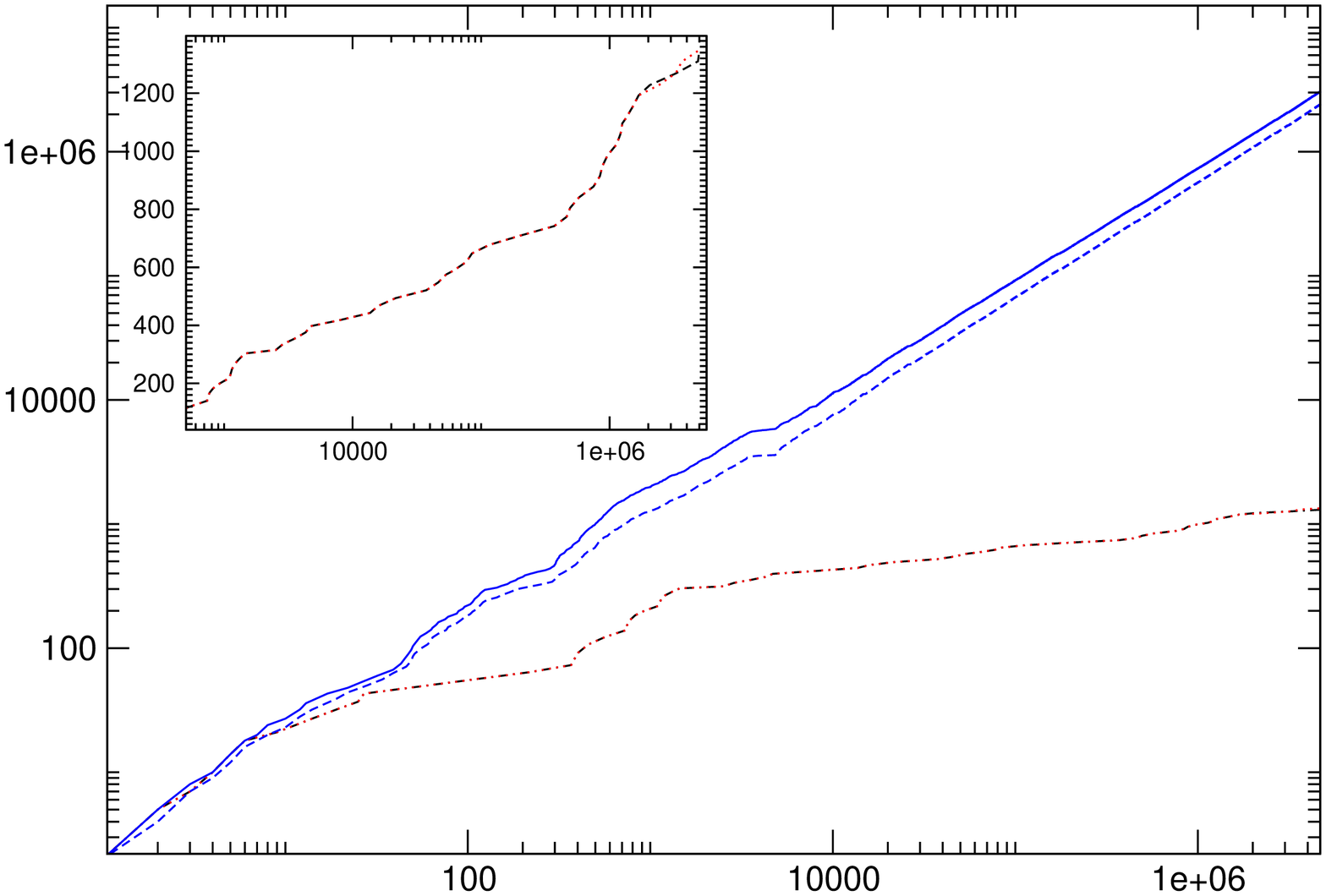,width=6.4cm,angle=270}}} 
&{\raggedleft{\psfig{figure=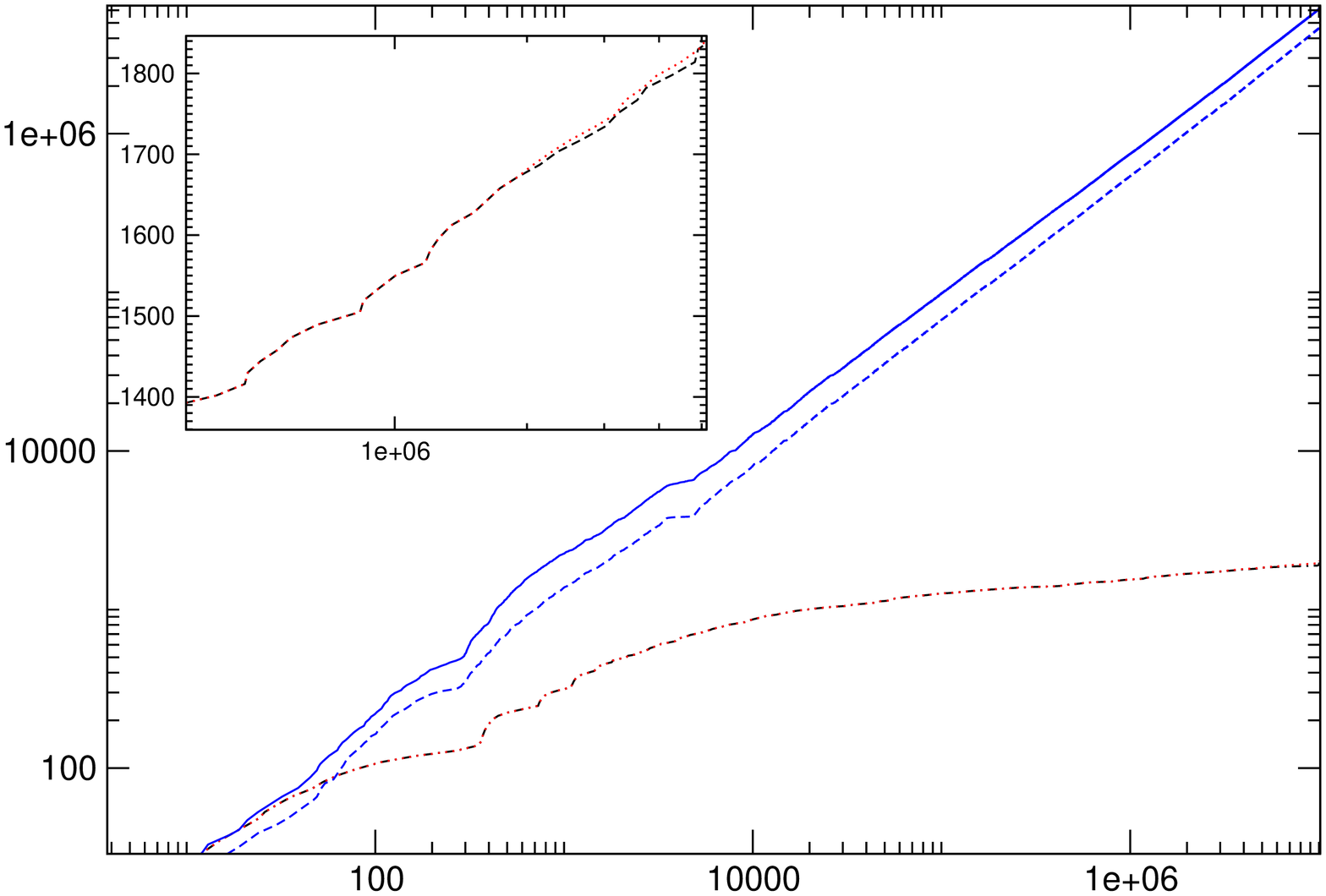,width=6.4cm,angle=270}}} 
\end{tabular} 
\caption{\it Information content for the Froeschl\'e map. On the left 
the results are obtained using CASToRe, on the right using LZ77. In 
both the pictures, the solid and dashed lines are for the full chaotic orbit, and 
the dotted curves are for the regular orbit. In the insets only the 
regular orbit is plotted in a log-linear scale.} 
\label{figfroe} 
\end{figure}

This map was studied in \cite{froe} as an example of a symplectic map 
for which it is possible to find the integrable and non-integrable 
initial conditions. The map is defined on the two-torus 
$\mathbb{R}^2/(2\pi \mathbb{Z})^2$ by 
$$\begin{array}{ll} 
x_{n+1} = & x_n +a \sin y_n \\ 
y_{n+1} = & x_n + y_n + a \sin y_n 
\end{array}$$ 
with $a=1.3$. We studied the behavior of the information content for 
orbits generated by two different initial conditions, one 
corresponding to the regular zone, and the other to the irregular 
zone. Both orbits have been studied with CASToRe and with LZ77.  In 
the regular zone, the initial condition is given by the point 
$(0,2.5)$. From the results in figure \ref{figfroe} (dotted curves), one 
can see that both the compression algorithms give indication of 
regularity by an increasing of the information content of the order of 
a logarithm (this behavior is visible clearly in the inset in the two 
figures, where the information content is plotted on a log-linear 
scale). The two compression algorithms also give a strong indication 
of full chaos for the irregular orbit, generated by the initial 
condition $(2,0)$. In figure \ref{figfroe}, the information content is 
a straight line (solid and dotted lines, with partitions in two vertical strips 
and in four equal squares) whose angular coefficients give an 
indication of the value 0.40 and 0.44, respectively, of the 
Kolmogorov-Sinai entropy with respect to the measure associated to the 
initial point. These results are in agreement with those of 
\cite{cohen-pro}. 
 
\end{document}